\documentclass[journal]{IEEEtran}
\IEEEoverridecommandlockouts
\usepackage{cite}
\usepackage{color}
\usepackage{amsmath,amssymb,amsfonts}
\usepackage{amsthm}
\usepackage[ulem={normalem,normalbf}]{changes}
\setuptodonotes{fancyline}
\newtheorem{definition}{Definition}
\newtheorem{theorem}{Theorem}

\newtheorem{assumption}{Assumption}

\newenvironment*{prooff}{\noindent \textbf{Proof}.}{\hfill $\blacksquare$ \vskip 4mm}
\usepackage{stfloats}
\usepackage[lined, ruled]{algorithm2e}
\usepackage{algpseudocode}
\SetKwProg{Init}{Initialization:}{}{}
\SetKwRepeat{Repeat}{}{until}
\SetKwRepeat{lRepeat}{repeat}{}
\usepackage{graphicx}
\usepackage{caption}
\usepackage{subcaption}
\usepackage{textcomp}
\usepackage{xcolor}
\usepackage{mathtools}
\usepackage{array}
\usepackage{bm}
\usepackage{booktabs}
\def\BibTeX{{\rm B\kern-.05em{\sc i\kern-.025em b}\kern-.08em
		T\kern-.1667em\lower.7ex\hbox{E}\kern-.125emX}}

\usepackage{hyperref}
\usepackage{xurl}
\usepackage{enumitem}
\begin{document}

\title{Selfish Cooperation Towards Low-Altitude Economy: Integrated Multi-Service Deployment with Resilient Federated Reinforcement Learning}

\author{Yuxuan Yang, Bin Lyu, \textit{Senior Member,~IEEE}, and Abbas Jamalipour, \textit{Fellow,~IEEE}
\thanks{Yuxuan Yang and Abbas Jamalipour are with the School of Electrical and Computer Engineering, University of Sydney, Sydney,
NSW, 2006 Australia (e-mail: yuxuan.yang@sydney.edu.au; a.jamalipour@ieee.org).}
\thanks{Bin Lyu is with the School of Communications and Information Engineering, Nanjing University of Posts and Telecommunications, Nanjing, 210003, China (e-mail: blyu@njupt.edu.cn).}
}

\maketitle

\begin{abstract}
The low-altitude economy (LAE) is a rapidly emerging paradigm that builds a service-centric economic ecosystem through large-scale and sustainable uncrewed aerial vehicle (UAV)-enabled service provisioning, reflecting the transition of the 6G era from technological advancement toward commercial deployment. The significant market potential of LAE attracts an increasing number of service providers (SPs), resulting in intensified competition in service deployment. In this paper, we study a realistic LAE scenario in which multiple SPs dynamically deploy UAVs to deliver multiple services to user hotspots, aiming to jointly optimize communication and computation resource allocation. To resolve deployment competition among SPs, an authenticity-guaranteed auction mechanism is designed, and game-theoretic analysis is conducted to establish the solvability of the proposed resource allocation problem. Furthermore, a resilient federated reinforcement learning (FRL)–based solution is developed with strong fault tolerance, effectively countering transmission errors and malicious competition while facilitating potential cooperation among self-interested SPs. Simulation results demonstrate that the proposed approach significantly improves service performance and robustness compared with baseline methods, providing a practical and scalable solution for competitive LAE service deployment.
\end{abstract}

\maketitle
\section{Introduction}\label{intro}




\IEEEPARstart{A}{s} mobile communications evolve from the fifth generation (5G) to the sixth generation (6G), many disruptive paradigms are transitioning from conceptual to practical, shifting the focus from technology-centric key performance indicators (KPIs), such as latency and energy consumption, to service-oriented key value indicators (KVIs), such as service quality and resource allocation efficiency \cite{9861699}. Among all the 6G enablers, unmanned aerial vehicle (UAV)-assisted wireless networks have attracted significant attention \cite{9768113, Jin2026}. 

The current decade has seen rapid growth in UAV adoption across public and industrial sectors, where UAVs can serve as aerial base stations (BSs), relays, and edge servers due to their reliable line-of-sight (LoS) links and flexible mobility. These expanding applications have fostered a new economic paradigm, referred to as the low-altitude economy (LAE) \cite{PernWernerSchmitXiao2024, 11259061}. According to industry white papers \cite{ZTE_LowAltitude_ISAC_2024}, the digital economy has become a cornerstone of global economic development, and LAE is regarded as one of its most promising segments, with widespread commercial applications such as express delivery, emergency response, and traffic monitoring.



As the market potential of LAE expands, diverse paradigms have emerged, including integrated sensing and communication (ISAC)-enabled LAE \cite{10879807,11072035}, LAE under multiple-input–multiple-output (MIMO) cellular systems \cite{10759668}, and post-disaster communications in LAE \cite{11051254}. However, the growing system complexity makes conventional optimization methods inadequate for efficient LAE service deployment. 

The rapid evolution of artificial intelligence (AI) has facilitated the development of agentic LAE operations, in which autonomous agents can effectively learn the complexity and dynamics of the environment to enable reliable service deployment \cite{11370843,9354996}. Among various AI techniques, deep reinforcement learning (DRL) has emerged as a powerful paradigm for LAE service deployment. Through centralized training, DRL-based solutions are capable of learning optimal policies in high-dimensional and non-linear environments, thereby enabling the integrated optimization of communication and computing resources across ground and aerial layers to support diverse LAE services \cite{wu2024deep}.

Achieving ubiquitous LAE service access remains a distant goal despite advances in 5G and 6G technologies. Many regions, especially sparsely populated rural areas, still lack basic data services \cite{dang2020should}. Temporary network access, such as post-disaster recovery and polar expeditions, demands both latency-sensitive and computation-intensive services \cite{9832657}. These conditions make centralized-training DRL inefficient due to excessive communication overhead and privacy risks from transmitting large volumes of raw data. Consequently, a communication-efficient and fault-tolerant architecture is essential for reliable LAE service deployment, particularly in infrastructure-limited scenarios.

Federated reinforcement learning (FRL) adopts a model-to-data paradigm, where each agent trains its policy locally and exchanges only model parameters with a global server, thereby reducing communication overhead and privacy risks \cite{zhuo2019federated}. This local–global architecture is well-suited to LAE applications, where UAVs can easily form hierarchical structures. A group of UAVs act as agents, and a higher altitude UAV or even low-earth-orbit (LEO) satellite serves as the global server, constructing space-air-ground integrated networks (SAGINs) to enable large-scale service deployment \cite{10745905}. 

However, most existing FRL-based solutions implicitly assume negligible inter-agent interference, which rarely holds in practical LAE scenarios where multiple service providers (SPs) deploy multiple services over shared hotspots, leading to inevitable competition. Moreover, studies on competition among multiple SPs in LAE are still at an early stage, despite being crucial for the rapidly expanding LAE market with increasing SPs entering. According to the Expert Market Research report, the Australian UAV market, valued at AUD 773.02 million in 2025, is experiencing rapid growth, accompanied by an increasingly competitive landscape \cite{EMR_Australia_Drone_Market_2026_2035}.



\begin{figure*}[ht]
		\centering
		\includegraphics[scale=0.4]{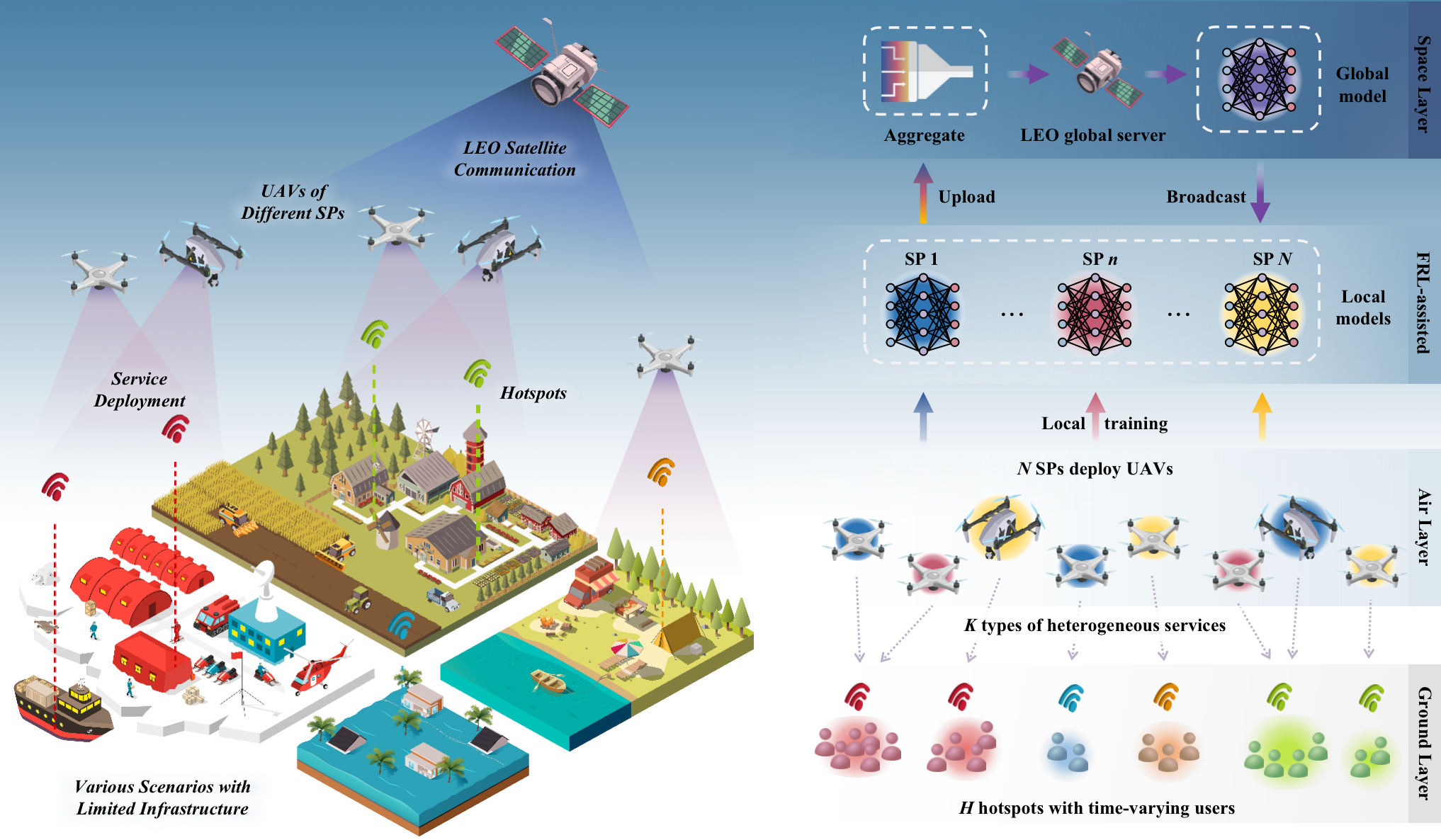}
		\renewcommand{\figurename}{Fig.}
		\caption{Multi-SP LAE system model and FRL-based framework.}
		\label{ssm}
	\end{figure*}





Therefore, this study is motivated by the need for a realistic and fault-tolerant framework that captures multi–SP competition and enables efficient and reliable service deployment for a wide range of LAE applications. As illustrated in Fig. \ref{ssm}, we consider pubulic service or infrastructure-limited scenarios such as rural areas, polar expeditions, and disaster-stricken regions, where sparsely distributed ground users form multiple hotspots. Multiple SPs deploy UAVs over these hotspots to provide multiple services by allocating communication and computing resources.

To achieve efficient cross-hotspot multi-service deployment, we design an authenticity-guaranteed auction mechanism to resolve competition among SPs within each hotspot. Building on this mechanism, an FRL-based framework is developed, where each SP acts as an agent and learns local resource allocation policies based on hotspot states, while a third-party global server (e.g., a LEO satellite) aggregates and updates the policy models. Moreover, a fault-tolerant mechanism is incorporated to maintain solution effectiveness even in the presence of transmission errors or malicious SP behaviors.

The contributions of this study are summarized as follows:
    
\begin{itemize}
    \item A realistic LAE resource allocation framework is proposed, in which multiple SPs dynamically deploy UAVs and compete to deliver diverse services across hotspots. The long-term utility of each SP is formulated as the optimization objective.
    

    \item An authentic auction mechanism is introduced to manage the inter-SP competition. Game-theoretical analysis proves that the formulated optimization problem can be solved by reaching the Nash Equilibrium (NE) set.

    \item We propose a resilient FRL-based solution that enables efficient integrated resource allocation for each SP by jointly designing real and virtual auctions. In addition, a dynamic filtering mechanism is introduced to tolerate transmission errors and malicious behaviors.
    

    \item Theoretical analysis and simulations verify the effectiveness, adaptability, and robustness of the proposed method, which establishes a scalable FRL-based paradigm for competitive multi-agent systems in LAE.
\end{itemize}

The rest of this paper is organized as follows: Related works are summarized in Section \ref{RW}. The system model is formulated in Section \ref{sm}. Section \ref{GM} presents an analysis of the multi-SP auction. The proposed solution is detailed in Section \ref{solution}, while Section \ref{PE} evaluates its performance. Finally, conclusions are drawn in Section \ref{conclusion}.

\section{Related Work}\label{RW}


The promising market potential of LAE has attracted increasing research attention. In \cite{10879807}, the authors studied a ground BS-supported LAE and proposed an alternating successive convex approximation (SCA) method to jointly optimize BS transmit beamforming, UAV trajectories, and BS associations. An LAE-oriented ISAC system was investigated in \cite{11072035}, where a DRL–based framework was developed to jointly optimize ground BS beamforming and UAV trajectories. Li et al. proposed a semi-definite relaxation (SDR)–based beamforming design for LAE MIMO cellular systems to simultaneously serve legitimate users and counter unauthorized UAVs \cite{10759668}. Zheng et al. focused on post-disaster communications in LAE and developed a diffusion model–enabled particle swarm optimization approach to maximize transmission rates through collaborative UAV beamforming \cite{11051254}.



DRL-based network optimization has been extensively explored. In \cite{wu2024deep}, the authors investigated a reconfigurable intelligent surface (RIS)-mounted UAV relay network and proposed an extended long short-term memory (LSTM)-double deep Q-network (DDQN) framework to jointly optimize UAV trajectory, RIS phase shift, and active transmit beamforming matrix. In \cite{zhang2023multi}, a multi-agent deep deterministic policy gradient (MADDPG)-based task scheduling optimization approach was presented, maximizing the number of offloading tasks for UAV energy efficiency. Tang et al. studied a rescue operation framework, which involves UAVs, ground-embedded robots, and airships, and proposed a MADDPG-based method for exploration and task assignment decisions optimization \cite{11106256} . Xie et al. explored RIS-mounted UAV for millimeter wave (mmWave) communications in LAE, and developed a DRL-based method with LSTM to jointly optimize RIS phase shifts and UAV trajectory \cite{11130648}.


FRL has also become a research focus due to its potential of privacy preservation and system overhead reduction compare to centralized DRL. In \cite{karmakar2023novel}, the authors proposed an FRL solution to ensure fairness by jointly optimizing UAV 3D trajectories, transmission power, and scheduling time for task offloading. Wu et al. studied an MADDPG-based FRL and developed a trajectory optimization method for emergency BSs to maximize average spectrum efficiency \cite{wu2022distributed}. In \cite{xu2023joint}, a dual-layer UAV-assisted network was examined, where a personalized FRL approach was introduced to enhance long-term network throughput and ensure user privacy. Zhang et al. investigated UAV-enabled data collection and proposed a FRL–based approach to collaboratively optimize UAV trajectories, scheduling, and completion time for minimizing the average age of information \cite{10980172}.



Although several service deployment paradigms have been proposed for LAE, existing studies (e.g., \cite{10879807, 11072035, 10759668, 11051254}) typically assume a single SP, which fails to capture the increasingly competitive LAE market \cite{EMR_Australia_Drone_Market_2026_2035}. DRL–based solutions \cite{wu2024deep, zhang2023multi, 11106256, 11130648}, particularly MADDPG, rely on centralized training, while such a design incurs excessive communication overhead and suffers from poor robustness in infrastructure-limited environments. FRL approaches \cite{karmakar2023novel, wu2022distributed, xu2023joint, 10980172} commonly assume seamless cooperation among agents, overlooking practical challenges such as inter-agent competition. Also, fault tolerance has not been addressed properly. 

In summary, to the best of our knowledge, this work is the first to develop a fault-tolerant FRL–based resource allocation framework for multi–SP competition, providing a robust solution for multi-service deployment in infrastructure-limited LAE scenarios while managing competition and enabling implicit cooperation among SPs with theoretical guarantees.


\section{System Model}\label{sm}
As shown in Fig. \ref{ssm}, in the considered system, $N$ SPs provide $K$ types of services to $H$ ground user hotspots by deploying UAVs over a given time period $T$. At each time $t \in \mathcal{T} = \{1\ldots T\}$, for each service type $k \in \mathcal{K} = \{1, \ldots, K\}$,  each SP $n \in \mathcal{N} = \{1\ldots N\}$ deploys a group of UAVs to support service provisioning. Each UAV is assumed to be equipped with a normalized unit of computing capability and communication bandwidth. Accordingly, an SP can realize dynamic resource deployment by flexibly adjusting the number of UAVs assigned to a given service and hotspot.

To simplify the analysis, the UAV swarm corresponding to each service type is abstracted as a logical UAV. Since this work focuses on service deployment and resource allocation rather than physical UAV scheduling, such an abstraction is reasonable and widely adopted at the decision-making level in UAV swarm studies, where coordinated behaviors and scheduling of multiple UAVs are learned and executed through centralized or shared decision-making mechanisms \cite{10833787,10946855}.

The number of active users $m_{h}(t) \triangleq |\mathcal{M}_{ht}|$ in hotspot $h \in \mathcal{H} = \{1\ldots H\}$ at time $t$ varies dynamically, where each user $i \in \mathcal{M}_{ht}$ has a task to be served. To provide global coordination, a LEO satellite functions as the global server, collecting, filtering and aggregating trustworthy local information from all SPs\footnote{The detailed methodology is provided in Section \ref{solution}.}.

\subsection{Computing Model}

For user $i$ in hotspot $h$, its task at time $t$ can be expressed as a tuple $\Gamma_{hi}(t)\triangleq\{D_{hi}(t),\lambda_{hi}(t),\{\chi_{hik}(t)\}_{k\in\mathcal{K}}\}$, where $D_{hi}(t)$ is the data size of the task (bits), $\lambda_{hi}(t)$ is the required capability of the task (CPU cycles per bit), and $\chi_{hik}(t)$ is the binary preference for service type $k$. If $\chi_{hik}(t)=1$, then user $i$ in hotspot $h$ requires type service $k$ at time $t$, otherwise $\chi_{hik}(t)=0$. The preference satisfies:

\begin{equation}
\sum_{k\in\mathcal{K}}\chi_{hik}(t)=1, \forall,h,i,t\in\mathcal{H},\mathcal{M}_{ht},\mathcal{T},
\end{equation}
which means that at each time $t$, user $i$ in hotspot $h$ can only requires for one type of service. Therefore, the overall required CPU cycles $\Lambda_{hk}(t)$ of hotspot $h$ of service $k$ at time $t$ can thus be calculated by:

\begin{equation}
    \Lambda_{hk}(t) = \sum_{i\in\mathcal{M}_{ht}} D_{hi}(t) \lambda_{hi}(t) \chi_{hik}(t).
\end{equation}

Each SP deploys $K$ UAVs equipped with computing capability over each hotspot to provide corresponding service to ground users. The computational capability allocated from SP $k$ to hotspot $h$ at time $t$ can be expressed as $F_{nhk}(t)$. For economic considerations, each SP $n$ is subject to a resource budget for service type $k$, denoted by $F_{nk}^{\max}$, i.e., $\sum_{h\in\mathcal{H}}F_{nhk}(t) \leq F_{nk}^{\max}$, $\forall n\in\mathcal{N}$ ,$ k\in\mathcal{K}$,  $t\in\mathcal{T}$. It is worth noting that $F_{nk}^{\max}$	represents a deployment budget rather than a physical resource limit. Each SP typically possesses resources exceeding this budget. Thus, this constraint does not imply physical resource reallocation across different hotspots.

Therefore, the total processing delay of SP $n$ providing service $k$ in hotspot $h$ at time $t$ is defined by:
\begin{equation}\label{T_comp}
    T_{nhk}^{\textup{Comp}}(t) = \frac{\Lambda_{hk}(t)}{F_{nhk}(t)}.
\end{equation}

\subsection{Communication Model}
Similarly, each SP is subject to a communication resource budget for service type $k$, denoted by $B_{nk}^{\max}$. When SP $n$ deploys UAVs to serve hotspot $h$, bandwidth $B_{nhk}(t)$ is allocated to the UAVs for communication at time $t$. Due to the geographical separation among different hotspots, inter-hotspot interference can be neglected. Within each hotspot, time-division multiple access is adopted for data transmission between the serving UAVs and ground users. Accordingly, the total communication delay $T_{nhk}^{\textup{Comm}}(t)$ is defined as the sum of the transmission times of user data, given by:
    \begin{equation}\label{T_comm}
        T_{nhk}^{\textup{Comm}}(t) = \frac{1}{R_{nhk}(t)}\sum_{i\in\mathcal{M}_{ht}}D_{hi}(t)\chi_{hik}(t),
    \end{equation}
    where $R_{nhk}(t)$ denotes the average transmission rate from hotspot $h$ to UAV $k$ of SP $n$.
    

The channel between UAVs and ground users is modeled as a statistical LoS channel, where the link may be in either LoS or non-LoS (NLoS) conditions with certain probabilities due to obstructions. Moreover, this work focuses on system-level resource allocation rather than instantaneous link reliability. As a result, fast fading on the order of milliseconds has a negligible impact on the overall performance. Inspired by \cite{8672604}, only large-scale fading is considered. Accordingly, the average transmission rate is expressed as
\begin{equation}
    R_{nhk}(t) = B_{nhk}(t)\log_2\left(1+\frac{\bar{p}_{h}(t)\,10^{-\frac{L_{\mathrm{dB}}}{10}}}{\sigma^2}\right),
\end{equation}
where $\bar{p}_{h}(t)$ is the average transmit power of users in hotspot $h$ at time $t$, $\sigma^2$ is the power of the Gaussian noise, and $L_{\mathrm{dB}}$ denotes the path loss between UAV and ground users. Users within the same hotspot are abstracted as a single entity, referred to as hotspot $h$.

Similar to \cite{8672604}, the path loss $L_{\mathrm{dB}}$ is given by
\begin{equation}
\begin{split}
    L_{\mathrm{dB}} 
    &= 20\log_{10}\!\left(\frac{4\pi f_c d_{nhk}(t)}{c}\right) \\
    &\quad + P_{\mathrm{LoS}}\eta_{\mathrm{LoS}} 
    + (1 - P_{\mathrm{LoS}})\eta_{\mathrm{NLoS}},
\end{split}
\end{equation}
where $f_{c}$ is the carrier frequency (Hz) and $c$ denotes the speed of light (m/s). $d_{nhk}(t)$ is the UAV altitude (m). $\eta_{\mathrm{LoS}}$ and $\eta_{\mathrm{NLoS}}$ denote the additional loss for LoS and NLoS links, respectively. $P_{\mathrm{LoS}}$ represents the probability of the channel being a LoS link and is calculated as:
\begin{equation}
    P_{\mathrm{LoS}} = \frac{1}{1+\alpha\exp(-\beta(\frac{\pi}{2}-\alpha))},
\end{equation}
where $\alpha$ and $\beta$ are environment-dependent variables\cite{7510820}.

Regarding satellite communications, satellites in the considered system only serve as a global coordinator in the FRL framework for model aggregation, rather than participating in data transmission for service provisioning. Compared with task-related data, the size of model parameters exchanged during FRL is relatively small, and the corresponding communication overhead can be neglected. Therefore, this work focuses on the communication links between UAVs and ground users, while omitting an explicit satellite communication model.

\subsection{Auction Mechanism}\label{Auction}
To address the competition that arises when multiple SPs provide the same service at a hotspot simultaneously, we introduce an auction mechanism. Specifically, the bid submitted by each SP for service $k$ at hotspot $h$ is its resource allocation pair, i.e., $\{F_{nhk}(t), B_{nhk}(t)\}$. From the perspective of service quality, the primary concern is the total processing delay. Accordingly, the total processing delay derived from the SP's allocation pair is expressed as:
    \begin{equation}
        T_{nhk}(t) = T_{nhk}^{\textup{Comm}}(t) + T_{nhk}^{\textup{Comp}}(t).
    \end{equation}

An SP may not deploy the actual resources exactly as declared in its bid, particularly when overbidding is adopted to increase the chance of winning the auction. We distinguish between the committed and actual processing delays. Specifically, $\hat{T}_{nhk}(t)$ denotes the committed processing delay implied by the submitted resource allocation (bid), while $T_{nhk}(t)$ represents the actual processing delay determined by the realized resource deployment. The auction outcome is determined based on the committed processing delay $\hat{T}_{nhk}(t)$, and the winner at time $t$ is determined by:
    \begin{equation}
        n^{*} = \mathop{\arg\min}\limits_{n\in\mathcal{N}}\hat{T}_{nhk}(t).
    \end{equation}

A binary variable is defined to indicate whether SP $n$ wins the auction of service type $k$ at hotspot $h$, given by:
    \begin{equation}
        \begin{split}
            W_{nhk}(t)= \left\{ \begin{array}{ll}
                1, & \textup{If} \ n = n^{*},\\
                0, & \textup{If} \ n \neq n^{*}.
            \end{array} \right.
        \end{split}
    \end{equation}
To identify overbidding, we introduce a verification indicator, denoted by $\hat{W}_{nhk}(t)$, defined as:
    \begin{equation}
        \hat{W}_{nhk}(t)=
        \begin{cases}
        1, & \text{if } T_{nhk}(t) \le \hat{T}_{nhk}(t),\\
        0, & \text{otherwise}.
        \end{cases}
    \end{equation}
    If $T_{nhk}(t) > \hat{T}_{nhk}(t)$, SP $n$ is considered to have overbid in the auction for service type $k$ at hotspot $h$. 

    Notably, such an auction mechanism operates without an auctioneer. UAVs from different SPs submit their bids directly to the hotspot, which autonomously selects the highest bid for data transmission without broadcasting the winner, thereby constituting a sealed bidding. This approach not only minimizes communication overhead at the hotspot but also prevents losing SPs from adjusting their bidding strategies based on the winner's actions, thereby maintaining the fairness of the auction. 

\subsection{Problem Formulation}\label{PF}
Based on the above models, the utility of each SP can be constructed. In LAE scenarios targeting resource-scarce regions or public services, the utility of an SP cannot be solely characterized by physical consumption. Otherwise, the dominant strategy would be to allocate no resources and provide no service. From a service deployment perspective, SPs compete for service provision rights of a specific service type at each hotspot. Consequently, regardless of the auction outcome, each SP incurs an opportunity cost, which is closely related to the total processing delay and may vary across different service types. 

We therefore model the opportunity cost as the product of the total processing delay and a type-dependent unit-time service cost coefficient $C_{nk}$. In addition, the auction winner receives a bonus $V_{nk}$ to further incentivize sustained deployment of service type $k$ at the hotspot. Accordingly, the SP utility is defined as:

\begin{equation}\label{EC_SP}
    U_{n}^{\textup{SP}}(t)= -\sum_{h\in\mathcal{H}}\sum_{k\in\mathcal{K}}[C_{nk}T_{nhk}(t)-V_{nk}W_{nhk}(t)\hat{W}_{nhk}(t)].
\end{equation}
It is worth noting that the cost is based on the actual processing delay, since the final service deployment relies on the realized resource allocation, regardless of the bidding strategy. The introduction of the indicator $\hat{W}_{nhk}(t)$ ensures that if an SP is identified as overbidding, the bonus will be revoked even when it wins the auction. This mechanism effectively discourages overbidding and encourages SPs to bid honestly.

This utility can flexibly switch between KPI-oriented and KVI-oriented evaluations, depending on the focus of specific LAE applications. For KVI-oriented scenarios, the cost coefficient and the bonus represent the economic cost and revenue of providing service type $k$, respectively. For KPI-oriented scenarios, an energy consumption interpretation can be adopted, where the cost coefficient denotes the energy consumption per unit processing time and the bonus represents the energy replenished during a recharging opportunity.

Under the sealed-bid defined in Subsection \ref{Auction}, only the winning SP receives a handshake signal from the hotspot. Upon service completion, the corresponding UAV is released and allowed to return for recharging, which motivates the above bonus definition\footnote{The energy consumption setting is adopted in Section \ref{PE} for simulations.}. To maximize its own long-term utility, each SP for $n \in \mathcal{N}$ independently solves the following optimization problem:
    \begin{equation} \label{P1}
	\begin{split}
		& \max_{F_{nhk}(t),B_{nhk}(t)}\sum_{t \in \mathcal{T}}U_{n}^{\textup{SP}}(t), \\
		\textup{s.t.}\ &\textup{C1}:\sum_{h \in \mathcal{H}}F_{nhk}(t)\leq F_{nk}^{\textup{max}}, \\
            &\textup{C2}:\sum_{h \in \mathcal{H}}B_{nhk}(t)\leq B_{nk}^{\textup{max}},\\
            &\textup{C3}:\sum_{k\in\mathcal{K}}\chi_{hik}(t)=1.
	\end{split}
    \end{equation}

Each SP aims to optimize its bids, i.e., resource allocation pairs $\{F_{nhk}(t), B_{nhk}(t)\}$ for all service types across all hotspots, thereby maximizing its long-term utility over the entire time period $T$. The constraints include: C1 and C2, ensuring that the total computing and communication resources allocated for service type $k$ across all hotspots do not exceed the deployment budgets; and C3, ensuring that each ground user at any time $t$ can select only one service type.

Solving problem \eqref{P1} is challenging due to several key factors, including the dynamic nature of user demands and service preferences, competitive interactions among SPs under the sealed-bid auction mechanism, the non-convexity introduced by binary auction outcomes, and the increased dimensions arising from multiple service types and hotspots. Since each SP’s allocation decisions depend on the actions of competing SPs, it is essential to characterize their strategic interactions. To this end, we begin by analyzing the multi-SP auction game.

\section{Authenticity Guaranteed Game}\label{GM}
In this section, we construct and analyze an auction-based multi-SP game to provide theoretical support for solving problem \eqref{P1}. We first examine the authenticity of the auction mechanism introduced in Subsection~\ref{Auction}. Then, an equivalent reformulated auction with an associated utility representation is developed, based on which a multi-SP game is constructed. By analyzing the equivalent multi-SP game, we establish the existence of NE and show that problem \eqref{P1} can be solved by characterizing the NE set. Finally, to facilitate an FRL-based solution, the multi-SP game is reformulated as a Markov game, enabling policy learning under partial information. Each SP optimizes its resource allocation based on local observations, while still benefiting from global knowledge through the FRL setting.

\subsection{Authenticity \& Equivalent Reformulation}

In order to ensure the effectiveness of the auction mechanism, it is necessary to ensure its authenticity, which is defined as follows:

\begin{definition}[Authenticity of auction]
    In an auction, if all participants cannot obtain higher utility through false bids, the auction is authentic.
\end{definition}
The following theorem is based on the definition.
\begin{theorem}\label{AUTH_AUC}
    The auction mechanism proposed in Subsection \ref{Auction} is an authentic auction, where no SP can obtian higher utility via false bids.
\end{theorem}

\begin{prooff}
    Please refer to Appendix \ref{appA}. 
\end{prooff}

Building on the authenticity guaranteed in Theorem \ref{AUTH_AUC}, we assume all SPs adopt authentic bidding. The auctions at each time $t$ are independent, as users in hotspots have time-varying service preferences and task requirements, leading to minimal correlation between auctions at different times. Consequently, the analysis focuses on the auction at a specific time $t$, without considering previous results as additional information.

The bidding action of SP $n$ at time $t$ is defined as a set $a_{n}(t) = \{F_{nhk}(t), B_{nhk}(t)\}_{h\in\mathcal{H},k\in\mathcal{K}}$, representing $HK$ simultaneous auctions for each SP. To facilitate theoretical analysis, we introduce an equivalent reformulation of the original auction. Specifically, the bonus granted to the winner is transformed into a penalty imposed on all non-winning SPs. For each SP, the cost of providing service type $k$ at hotspot $h$ is redefined as:
    \begin{equation}\label{Mod_C}
        \begin{split}
            C_{nhk}(t) = \left\{ \begin{array}{ll} 
            C_{nk}T_{nhk}(t) & \textup{If} \ W_{nhk}(t) = 1,\\
            C_{nk}T^{\textup{Pen}} & \textup{If} \ \ W_{nhk}(t) = 0,
            \end{array} \right.
        \end{split}
    \end{equation}
where $T^{\textup{Pen}}\gg T_{nhk}(t)$ is the time penalty. Based on this, the utility of SP is redefined as:

\begin{equation}\label{mod_U}
    \bar{U}_{n}^{\textup{SP}}(t) = - \sum_{h\in\mathcal{H}}\sum_{k\in\mathcal{K}}C_{nhk}(t).
\end{equation}

The following theorem ensures that the redefined utility preserves all properties of the original auction:
\begin{theorem}\label{Theo:equivalent}
    The utility $U_{n}^{\textup{SP}}(t)$ defined by Eq. \eqref{EC_SP}) and the utility $\bar{U}_{n}^{\textup{SP}}(t)$ defined by Eq. \eqref{mod_U} are equivalent.
\end{theorem}
\begin{prooff}
    Please refer to Appendix \ref{appE}.
\end{prooff} 

\subsection{Auction Game Formulation \& NE Analysis}
The modified SP utility is strongly influenced by the auction result, since a failure brings an explicit penalty, which enhances the original auction. The equivalence between the enhanced auction and the original one is guaranteed by Theorem \ref{Theo:equivalent}. As a result, the NE analysis based on the modified utility $\bar{U}_{n}^{\textup{SP}}(t)$ applies to the original auction.

As discussed earlier, at each time $t$, SP $n$ simultaneously participates in $HK$ auctions. These $HK$ auctions can be viewed as a stage game $\mathcal{G}_{t}^{\textup{SP}}$, as all bidding actions are coupled through the resource budget constraints C1 and C2. The stage game $\mathcal{G}_{t}^{\textup{SP}}$ and its corresponding NE are defined accordingly.

\begin{definition}[Stage game $\mathcal{G}_{t}^{\textup{SP}}$]\label{SG}
    At each time $t$, auctions of SPs bidding for $K$ types of services at $H$ hotspots, constitute a stage game \begin{equation}
        \mathcal{G}_{t}^{\textup{SP}} \triangleq \{\mathcal{N}, \{a_{n}(t)\}_{n\in\mathcal{N}}, \{\bar{U}_{n}^{\textup{SP}}(t)\}_{n\in\mathcal{N}}\}.
    \end{equation}
\end{definition}

\begin{definition}[NE of game $\mathcal{G}_{t}^{\textup{SP}}$]\label{defi_NE}
    The bidding action set of SPs deploying $K$ services for $H$ hotspots at time $t$ $\textbf{a}^{*}(t) = \{a_{n}^{*}(t)\}_{n\in\mathcal{N}}$ is the NE of game $\mathcal{G}_{t}^{\textup{SP}}$ if and only if no SP can maximize its utility $\bar{U}_{n}^{\textup{SP}}(t)$ by unilaterally deviating, i.e.,
    \begin{equation}\label{ie_NE}
        \bar{U}_{n}^{\textup{SP}}(a_{n}^{*}(t), \textbf{a}_{-n}^{*}(t)) \ge \bar{U}_{n}^{\textup{SP}}(a_{n}(t), \textbf{a}_{-n}^{*}(t)), \forall n\in\mathcal{N}.
    \end{equation}
\end{definition}

In order to reach the NE of the game of SPs, the potential function is defined by potential game theory\cite{MONDERER1996124}.

\begin{definition}\label{potentialF}
    For a game, if there is a global function $f_{p}(a_{n},\textbf{a}_{-n})$ such that for all $a_{n}'\ne a_{n}$, the utility function $u_{n}(a_{n},\textbf{a}_{-n})$ satisfies
    \begin{equation}
    \begin{split}
        &\textup{if} :\  u_{n}(a_{n}',\textbf{a}_{-n}) \ge u_{n}(a_{n},\textbf{a}_{-n}),\\
        &\textup{then}: \  f_{p}(a_{n}',\textbf{a}_{-n}) \ge f_{p}(a_{n},\textbf{a}_{-n}).
    \end{split}
    \end{equation}
    Then $f_{p}(a_{n},\textbf{a}_{-n})$ is the potential function of the game.
\end{definition}

Based on Definition \ref{potentialF}, the following theorem can then be inferred. 

\begin{theorem}\label{at_least_1NE}
    Stage game $\mathcal{G}_{t}^{\textup{SP}}$ among SPs has at least one NE, whose weighted potential function is defined by:
    \begin{equation}
    \begin{split}
        &\Phi^{\textup{SP}}(t) =\\ & -\sum_{n\in\mathcal{N}}\sum_{h\in\mathcal{H}}\sum_{k\in\mathcal{K}}[W_{nhk}(t)T_{nhk}(t)+(1-W_{nhk}(t))T^{\textup{Pen}}].
    \end{split}
    \end{equation}
\end{theorem}
\begin{prooff}
    Please refer to Appendix \ref{appB}.
\end{prooff}

According to Theorems \ref{AUTH_AUC} and \ref{at_least_1NE}, we prove that the game among SPs is authentic and has at least one NE. Then we continue to analyze the relationship between this NE and the solution of problem \eqref{P1}. Based on the Definitions \ref{SG} and \ref{defi_NE}, we can infer Theorem \ref{P1_solution}.

\begin{theorem}\label{P1_solution}
    The set of NE of the stage game $\mathcal{G}_{t}^{\textup{SP}}$, $\forall t \in \mathcal{T}$ constitutes the solution of the proposed problem in \eqref{P1}, i.e.,
    \begin{equation}
        \{a_{n}^{*}(t)\}_{t\in \mathcal{T}} = \mathop{\arg\max}\limits_{a_{n}(t)}\sum_{t\in\mathcal{T}}U_{n}^{\textup{SP}}(t), \forall n \in \mathcal{N}.
    \end{equation}
\end{theorem}

\begin{prooff}
    Please refer to Appendix \ref{appC}.
\end{prooff}

Based on Theorems \ref{AUTH_AUC}–\ref{P1_solution}, problem \eqref{P1} admits a solution characterized by the NE set. However, the above analysis assumes complete global information. In practical LAE scenarios, time-varying user populations, task demands, and service preferences render direct computation of the NE for each stage game intractable using conventional game-theoretic approaches. Nevertheless, the NE analysis establishes the theoretical solvability of problem \eqref{P1}, thereby providing a basis for the proposed FRL-based solution in Section \ref{solution}.

\subsection{Markov Game Formulation}
Multi-SP Markov game can be defined by a tuple $\left\langle \mathcal{N},\mathcal{S},\mathcal{O},\mathcal{A},\mathcal{P},\mathbb{R},\gamma \right\rangle$:
\begin{itemize}
    \item $\mathcal{N}$ is the set of SPs.
    \item $\mathcal{S} \triangleq \{s(t) = (\mathcal{S}_{1}(t), \mathcal{S}_{2}(t))\}$, $t \in \mathcal{T}$, is the state set, where $\mathcal{S}_{1}(t)$ is the bidding information of all SPs, i.e., $\mathcal{S}_{1}(t) \triangleq \{(F_{nhk}(t), B_{nhk}(t))\}_{n,h,k\in \mathcal{N,H,K}}$, and $\mathcal{S}_{2}(t)$ is the task information of all users, i.e., $\mathcal{S}_{2}(t) \triangleq \{\Gamma_{hi}(t)\}_{h,i\in \mathcal{H},\mathcal{M}_{ht}}$.
    \item $\mathcal{O} \triangleq \{o_{n}(t)\}$ is the observation set. We assume each SP can only observe its own auction results and task information of all users, i.e., $o_{n}(t) = \{\mathcal{S}_{2}(t),\{W_{nhk}(t-1)\}_{h,k \in \mathcal{H,K}}\}$. The bidding information of other SPs is unknown to each SP.
    \item $\mathcal{A} \triangleq \{\textbf{a}(t)\}$, $\forall t \in \mathcal{T}$ is the action set of all SPs. As defined previously, $\textbf{a}(t) = \{a_{n}(t)\}_{n \in \mathcal{N}}$, and $a_{n}(t) = \{F_{nhk}(t), B_{nhk}(t)\}_{h\in\mathcal{H},k\in\mathcal{K}}$.
    \item $\mathcal{P}: \mathcal{S}\times \mathcal{A}\times \mathcal{S}\to [0,1]$ is the state transition probability distribution, where the probability of system state transition from $s(t)$ to $s(t+1)$ is noted by $P(s(t+1)|s(t),\textbf{a}(t))$.
    \item $r_{n}(t): \mathcal{S}\times \mathcal{A} \to [0,R]$ is the reward of each SP ($R >0$ is a constant), which is defined by $r_{n}(t) \triangleq U_{n}^{\textup{SP}}(t)$. Thus, the accumulated discounted reward of each SP can be defined by $\bar{R}_{n}(t) \triangleq \sum_{\tau = 0}^{t}\gamma^{\tau}r_{n}(t)$, where $\gamma$ is the discounted factor.

\end{itemize}
Each SP $n$ generates a trajectory of experiences defined as $\tau_{n} \triangleq \{ o_{n}(0), a_{n}(0), r_{n}(0), o_{n}(1), \ldots \}$, comprising sequences of observations, actions, and rewards over time. These trajectories serve as the training foundation for the reinforcement learning algorithm.

\section{Dual-auction potential-cooperation Resilient Federated Policy Gradient Solution}\label{solution}

\begin{figure*}[ht]
		\centering
		\includegraphics[scale=0.45]{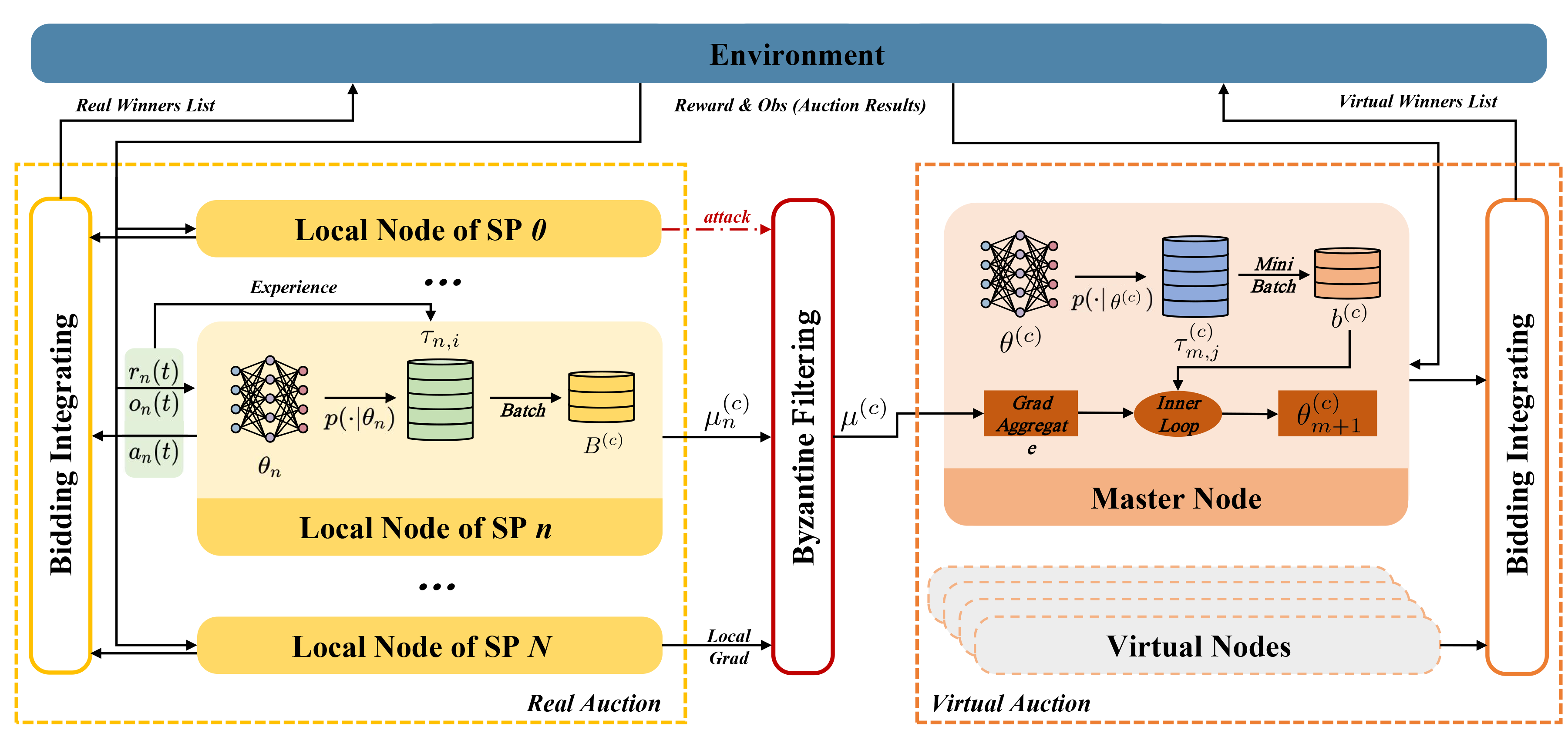}
		\renewcommand{\figurename}{Fig.}
		\caption{A training illustration of DAPCR-FedPG.}
		\label{DAPCR}
	\end{figure*}

This section proposes the resilient FRL solution, namely, Dual-Auction Potential-Cooperation Resilient Federated Policy Gradient (DAPCR-FedPG), where the LEO satellite functions as a global server (master node), and each SP maintains a local server (local node). As shown in Fig. \ref{DAPCR}, DAPCR-FedPG integrates three core components: real auction, virtual auction, and Byzantine filtering. Algorithm \ref{DA} presents the complete training procedure of DAPCR-FedPG. 

\begin{algorithm}[ht]
    \caption{Training of DAPCR-FedPG}\label{DA} 
    \textbf{Initialization:} 
    \begin{itemize}
        \item Policy parameters of all local nodes $\theta_{n}^{(0)},\forall n\in \mathcal{N}$,\\ and master node $\theta^{(0)}$
        \item Batch size $B^{(0)}$, and mini batch size $b^{(0)}$
        \item Learning rate $\eta$
    \end{itemize}
    \For {\textup{epoch} $c\in C$}{
    $\theta^{(c)}_{\mathrm{Re}},\theta_{n}^{(c)}
    \gets \theta^{(c-1)}$\\
        \For{\textup{all local nodes} $n\in\mathcal{N}$}{
        \begin{itemize}
            \item \textbf{Collecting real experience}\\
            Sample $B^{(c)}$ trajectories $\{\tau_{n,i}^{(c)}\}_{i=1}^{B^{(c)}}$ with \\ policy $\pi_{\theta_{n}^{(c)}}$
            \item \textbf{Local gradient estimation}\\
            Get $\mu_{n}^{(c)}$ based on Eq. \eqref{local_GE})
        \end{itemize}}
        Filtering out the good nodes set $\mathcal{Z}^{(c)}$ according to Algorithm \ref{DAPCR}\\
        $\mu^{(c)}=\frac{1}{|\mathcal{Z}^{(c)}|}\sum_{n\in\mathcal{Z}^{(c)}}\mu_{n}^{(c)}$\\
        Sample $M^{(c)}\sim Geom(\frac{B^{(c)}}{B^{(c)}+b^{(c)}})$\\
        \For{$m\in M^{(c)}$}{
        \begin{itemize}
            \item \textbf{Collecting virtual experience}\\
            Sample $b^{(c)}$ trajectories $\{\tau_{m,j}^{(c)}\}_{j=1}^{b^{(c)}}$ with \\ policy $\pi_{\theta_{m}^{(c)}}$
            \item \textbf{Master gradient estimation}\\
            Get $v_{m}^{(c)}$ based on Eq. \eqref{vt}
            \item \textbf{Parameters update}\\
            $\theta_{m+1}^{(c)} = \theta_{m}^{(c)} +\eta v_{m}^{(c)}$
        \end{itemize}}
        $\theta^{(c)} \gets \theta^{(c)}_{M^{(c)}}$
        }
\end{algorithm}

\subsection{Real Auction and Local Gradient Estimation}\label{RALGE}

The local node of SP $n$ with a local policy network generates bidding action $a_{n}(t)$ based on observation $o_{n}(t)$. After all nodes bid, hotspots transmit tasks to the winners. Following the real auction, each local node calculates the reward $r_{n}(t)$ and stores the auction result for the next round. The observation, action, and reward form an experience slice at time $t$, while experiences over the entire period $T$ constitute a trajectory, which is used to update local policy network parameters $\bm{\theta}_{n}$. The gradient is defined as follows:

\begin{equation}
    \nabla J(\bm{\theta}_{n}) = \mathbb{E}_{\tau_{n}\sim p(\cdot|\bm{\theta}_{n})}[\nabla \log p(\tau_{n}|\bm{\theta}_{n})\bar{R}_{n}(\tau_{n})],
\end{equation}
where $p(\tau_{n}|\bm{\theta}_{n})$ is the distribution of trajectory $\tau_{n}$. 

Sampling trajectories directly from $p(\tau_{n}|\bm{\theta}_{n})$ can result in the loss of gradient information. The GPOMDP gradient estimator \cite{baxter2001infinite} is applied to reduce variance while maintaining unbiasedness as it defines the accumulated reward as the return $\hat{R}_{n}(t)$ for gradient estimation, which is expressed as:

\begin{equation}
    \hat{R}_{n}(t) = \sum_{x=0}^{T-t}\gamma^{x}r_{n}(t+x). 
\end{equation}

To further minimize estimation error, the advantage value $A_{n}(t)$ is employed to assess the relative value of each action at time $t$, which is defined by:

\begin{equation}
    A_{n}(t) = \frac{\hat{R}_{n}(t) - \mu_{\hat{R}}}{\sigma_{\hat{R}}},
\end{equation}
where $\mu_{\hat{R}}$ and $\sigma_{\hat{R}}$ are the mean and standard deviation of the return $\hat{R}_{n}(t)$ over time period $T$. Thus, the gradient $g(\tau_{n}|\bm{\theta}_{n})$ with respect to the parameter $\bm{\theta}_{n}$, can be defined as follows:

\begin{equation}
    g(\tau_{n}|\bm{\theta}_{n}) = - \sum_{t\in \mathcal{T}} A(t) \cdot \nabla_{\bm{\theta}_{n}}\log \pi_{\bm{\theta}_{n}}(a_{n}(t)|o_{n}(t)).
\end{equation}

Considering the batch size of $B$, where each batch contains $B$ trajectories, the gradient estimate for the local node of SP $n$ is defined as follows:

\begin{equation}
    \widehat{\nabla}_{B} J(\bm{\theta}_{n}) = \frac{1}{B}\sum_{i=1}^{B}g(\tau_{n,i}|\bm{\theta}_{n}),
\end{equation}
where $\tau_{n,i}$ denotes the $i$-th trajectory in the batch.

\subsection{Byzantine Filtering with Dynamic Threshold}\label{DTBF}
After local gradient estimation, all nodes transmit their gradients to the master node for aggregation. Nodes transmitting correct gradients form the good-node set $n \in \mathcal{Z}$, while those sending false gradients are Byzantine nodes,  with proportion $\alpha_{\textup{B}}\triangleq\frac{|\mathcal{N}\setminus\mathcal{Z}|}{N}$. The gradient sent by a local node corresponding to each SP at epoch $c \in [0,C]$ is defined as:

\begin{equation}\label{local_GE}
    \mu_{n}^{(c)} =\left\{\begin{array}{ll}
    \hat{\nabla}J_{B^{(c)}}(\bm{\theta}_{n}^{(c)})\\=\frac{1}{B^{(c)}}\sum_{i=1}^{B^{(c)}}g(\tau_{n,i}|\bm{\theta}_{n}^{(c)})  & \textup{for} \ \ n \in \mathcal{Z}, \\
* & \textup{for} \ \ n \in \mathcal{N}\setminus\mathcal{Z},
\end{array} \right.
\end{equation}
where $*$ represents false gradient. To filter out Byzantine nodes, a gradient-based filtering mechanism is required, which relies on the following assumption:
\begin{assumption}\label{ass_on_grad_var}
    A constant $\varepsilon > 0$ exists for all policies $\pi_{\bm{\theta}}$, such that:
    \begin{equation}
        \textup{Var}(g(\tau|\bm{\theta})) \leq \varepsilon.
    \end{equation}
\end{assumption}

In stochastic non-convex optimization, Gaussian-based policies allow a fixed variance bound $\varepsilon$ . In contrast, the proposed policy network employs a Dirichlet distribution for policy construction, requiring a dynamic variance bound defined as
\begin{equation}
    \varepsilon^{(c)} = \textup{mean}(\left \| \mu_{n}^{(c)} - \mu_{n'}^{(c)} \right \|) + \omega_{\varepsilon}^{(c)}\textup{std}(\left \| \mu_{n}^{(c)} - \mu_{n'}^{(c)} \right \|), 
\end{equation}
for all $n,n' \in\mathcal{N}$, where $\omega_{\varepsilon}^{(c)}$ is a weight coefficient defined to weigh the proportion of standard deviation at epoch $c$. Inspired by \cite{khanduri2019byzantine}, the Dynamic Threshold Byzantine Filtering (DTBF) mechanism, detailed in Algorithm \ref{BF}, employs a two-stage process with stricter threshold $\mathfrak{T}_{\mu}^{(c)}$ and lenient threshold $2\varepsilon^{(c)}$, where $V \triangleq2\log \left(\frac{2N}{\delta}\right)$ and $\delta \in (0,1)$. 

Each filtering stage consists of three steps: (1) Identify the potential good gradient set $\tilde{\mathcal{N}}$ by evaluating pairwise distances among local gradients; (2) Determine the median gradient $\mu_{\textup{med}}^{(c)}$ as the one closest to the mean of $\tilde{ \mathfrak{U} }^{(c)}$; (3) Compare each local gradient’s distance to $\mu_{\textup{med}}^{(c)}$ to derive the good node set $\mathcal{Z}^{(c)}$.

\begin{algorithm}[t]
    \caption{Dynamic Threshold Byzantine Filtering}\label{BF} 
    \textbf{Input:}
    \begin{itemize}
        \item local gradient set $ \mathfrak{U}^{(c)} \triangleq\{\mu_{n}^{(c)}\}_{n\in\mathcal{N}}$
        \item dynamic variance bound $\varepsilon^{(c)}$
        \item gradient threshold $\mathfrak{T}_{\mu}^{(c)} \triangleq 2\varepsilon^{(c)} \sqrt{\frac{V}{B^{(c)}}}$, where $V \triangleq2\log \left(\frac{2N}{\delta}\right)$ and $\delta \in (0,1)$
    \end{itemize}
    \textbf{Initialization:} 
    \begin{itemize}
        \item good nodes set at epoch $c$, $\mathcal{Z}^{(c)} = \emptyset$
        \item threshold list $\mathcal{L}^{(c)} = [\mathfrak{T}_{\mu}^{(c)}, 2\varepsilon^{(c)}]$
    \end{itemize}
    \For {$\mathfrak{T}^{(c)}\in \mathcal{L}^{(c)}$}{
        \If{$|\mathcal{Z}^{(c)}| < (1-\alpha_{\textup{B}})N$}{
        \begin{itemize}
            \item \textbf{Identifying potential good gradients set}\\
            \[\hspace{-8em}
            \tilde{ \mathfrak{U}}^{(c)} \triangleq \left\{ \mu_{\tilde{n}}^{(c)} \right\}_{\tilde{n} \in \tilde{\mathcal{N}}}, \textup{where} \ \tilde{n} \in \tilde{\mathcal{N}} \subseteq \mathcal{N} \ \textup{s.t.}
            \]
            \[\hspace{-8em}
            \left|\left\{ n' \in \mathcal{N} : \left \| \mu_{n'}^{(c)} - \mu_{\tilde{n}}^{(c)} \right \| \leq \mathfrak{T}^{(c)} \right\}\right| > \frac{N}{2}
            \]
            \item \textbf{Finding median gradient}\\
            \[\hspace{-8em}
            \mu_{\textup{med}}^{(c)} \gets \arg\min_{\mu_{\tilde{n}}^{(c)} \in \tilde{ \mathfrak{U} }} \left \| \mu_{\tilde{n}}^{(c)} - \textup{mean}(\tilde{ \mathfrak{U} }^{(c)}) \right \|
            \]
            \item \textbf{Determining good nodes set}\\
            \[\hspace{-8em}
            \mathcal{Z}^{(c)} \triangleq \left\{ n \in \mathcal{N} : \left \| \mu_{n}^{(c)} - \mu_{\textup{med}}^{(c)} \right \| \leq \mathfrak{T}^{(c)} \right\}
            \]
        \end{itemize}
        }}
    \Return Good nodes set $\mathcal{Z}^{(c)}$
\end{algorithm}

\subsection{Virtual Auction and Parameters Update}\label{VAPU}
After identifying the good nodes set through DTBF, the master node aggregates their gradients as $\mu^{(c)}\triangleq\frac{1}{|\mathcal{Z}|}\sum_{n\in\mathcal{Z}}\mu_{n}^{(c)}$ and collects trajectories for gradient estimation. Typical FRL-based solutions assume identical yet independent environments \cite{fan2021fault} for all nodes, where each node's actions do not affect others' rewards. To address the competitive nature of our system, $N-1$ virtual nodes are introduced. Each virtual node inherits parameters $\theta^{(c)}_{\mathrm{Re}} \gets \theta^{(c-1)}$ from the previous epoch $c-1$, and gets master node's observations to form the virtual auctions, which essentially constitutes a self-update by competing with past policies.

To further reduce estimation variance, an inner loop with $M^{(c)}$ steps is introduced. The inner loop step size $M^{(c)}$ is sampled from the geometric distribution defined by $B^{(c)}$ and $b^{(c)}$. In each step $m \in M^{(c)}$, mini-batched trajectories $\{\tau_{m,j}^{(c)}\}_{j=1}^{b^{(c)}}$ of size $b^{(c)}\ll B^{(c)}$ are collected based on parameters $\theta^{(c)}_{m}$. Based on stochastic variance-reduced policy gradient (SVRPG) \cite{papini2018stochastic}, the master node's gradient estimation at step $m$ is defined as:
\begin{equation}\label{vt}
    v^{(c)}_{m} = \frac{1}{b^{(c)}}\sum_{j=1}^{b^{(c)}}[g(\tau_{m,j}^{(c)}|\theta_{m}^{(c)})-w_{m,j}^{(c)}g(\tau_{m,j}^{(c)}|\theta_{\textup{Re}}^{(c)})]+\mu^{(c)},
\end{equation}
where $w_{m,j}^{(c)} \triangleq p(\tau|\theta_{m}^{(c)})/p(\tau|\theta_{\textup{Re}}^{(c)})$ is the importance weight of the current gradient estimation $g(\tau_{m,j}^{(c)}|\theta_{m}^{(c)})$ relative to the previous one $g(\tau_{m,j}^{(c)}|\theta_{\textup{Re}}^{(c)})$. Master node then updates its parameters at step $m$ by:
\begin{equation}
    \theta_{m+1}^{(c)} = \theta_{m}^{(c)} + \eta v_{m}^{(c)},
\end{equation} 
where $\eta$ is the learning rate. When $m = M^{(c)}$, the parameters $\theta_{M^{(c)}}^{(c)}$ are broadcast to all local nodes, i.e., $\theta^{(c)} \gets \theta^{(c)}_{M^{(c)}}$.

As summarized in Algorithm \ref{DA}, after initializing the policy network and learning hyperparameters, local nodes participate in real auctions to estimate local gradients, which are aggregated by the master node using the DTBF algorithm, followed by virtual auctions with $N-1$ virtual nodes to update and broadcast the global model. 

Finally, during the execution phase, the complexity of the DAPCR-FedPG algorithm is $\mathcal{O}(T(\sum_{z=1}^{Z}n_{z}\cdot n_{z-1}))$, where $z\in Z$ represents the neural network layers, and $n_{z}$ denotes the number of neurons in the $z$-th layer.

\section{Performance Evaluation}\label{PE}

\subsection{Settings}\label{Settings}
The experimental setup involves 5 SPs providing 4 service types across 6 hotspots ($N = 5$, $H = 6$, $K = 4$), with 2 SPs acting as Byzantine nodes introducing significant noise into local gradient estimation. The time period considered is set to 30 ($T = 30$). Following  the configuration of RACHEL-Corrections 3.0 \cite{rachel_corrections_3}, the number of users per hotspot, $m_{h}(t)$, is chosen from the discrete set [20, 30, 40, 50].

The computing task parameters for each user, including data size $D_{hi}(t)$ and required CPU cycles per bit $\lambda_{hi}(t)$, are uniformly distributed within [2.5, 3, 3.5, 4, 4.5, 5] MB and [1000, 1200, 1400, 1600] CPU cycles/bit, respectively \cite{10147242}. The total resources for each service are set to $F_{nk}^{\textup{max}} = 800$ GHz and $B_{nk}^{\textup{max}} = 800$ MHz. The values $(\eta_{LoS}, \eta_{NLoS}, \alpha, \beta)$ are adopted from the suburban scenario in \cite{7510820}. 

In addition, an energy consumption interpretation of the utility is adopted. Specifically, the cost coefficient $C_{nk}$ is defined as the energy consumption per unit service time, calculated from the UAV power settings in \cite{9411725}. The bonus $V_{nk}$ for the auction winner is defined as an energy replenishment, corresponding to a short-duration docking recharge based on the DJI Matrice 3TD specification report in \cite{DJI_Dock2_Specs}. Specific parameter values are summarized in Table \ref{channel_para}. 

\begin{table}[h!]
\centering
\caption{Partial Model Parameter Settings}
\label{channel_para}
\resizebox{\columnwidth}{!}{%
\begin{tabular}{@{}llll@{}}
\toprule
\textbf{Parameter} & \textbf{Value} & \textbf{Parameter} & \textbf{Value} \\ 
\midrule
White Gaussian noise $\sigma^{2}$ & -95 dBm \cite{pan20193d} & 
Transmit power $\bar{p}_{h}(t)$ & 0.1 W \cite{10048752},\cite{9411725}\\
Carrier frequency $f_{c}$ & 2.5 GHz \cite{pan20193d},\cite{7510820} & Energy replenish $V_{nk}$ & $3000$ J (30s)\cite{DJI_Dock2_Specs}\\
Loss of LoS $\eta_{LoS}$ & 0.1 \cite{7510820} &
Loss of NLoS $\eta_{NLoS}$ & 21 \cite{7510820} \\
Environmental para. $\alpha$ & 4.88 \cite{7510820} & Environmental para. $\beta$ & 0.43 \cite{7510820}\\
Flight altitude $h_{nhk}(t)$ & 100 m \cite{10147242}, \cite{10048752} & Unit energy consumption $C_{nk}$ & 381 W \cite{9411725} \\
\bottomrule
\end{tabular}}
\end{table}

For the training parameters, both the local nodes and the master node employ a Dirichlet policy with three fully connected layers and the Adam optimizer. The batch size $B^{(c)}$ is uniformly distributed in [120, 130], with a mini-batch size $b^{(c)}$ fixed at 64. The discount factor is $\gamma = 0.9$, the learning rate is $\eta = 9\times 10^{-5}$ and 15,000 trajectories are collected.

\subsection{Performance Comparison}

\begin{figure}[ht]
\centering
    \begin{subfigure}{0.68\linewidth}
        \centering
        \includegraphics[width=\linewidth]{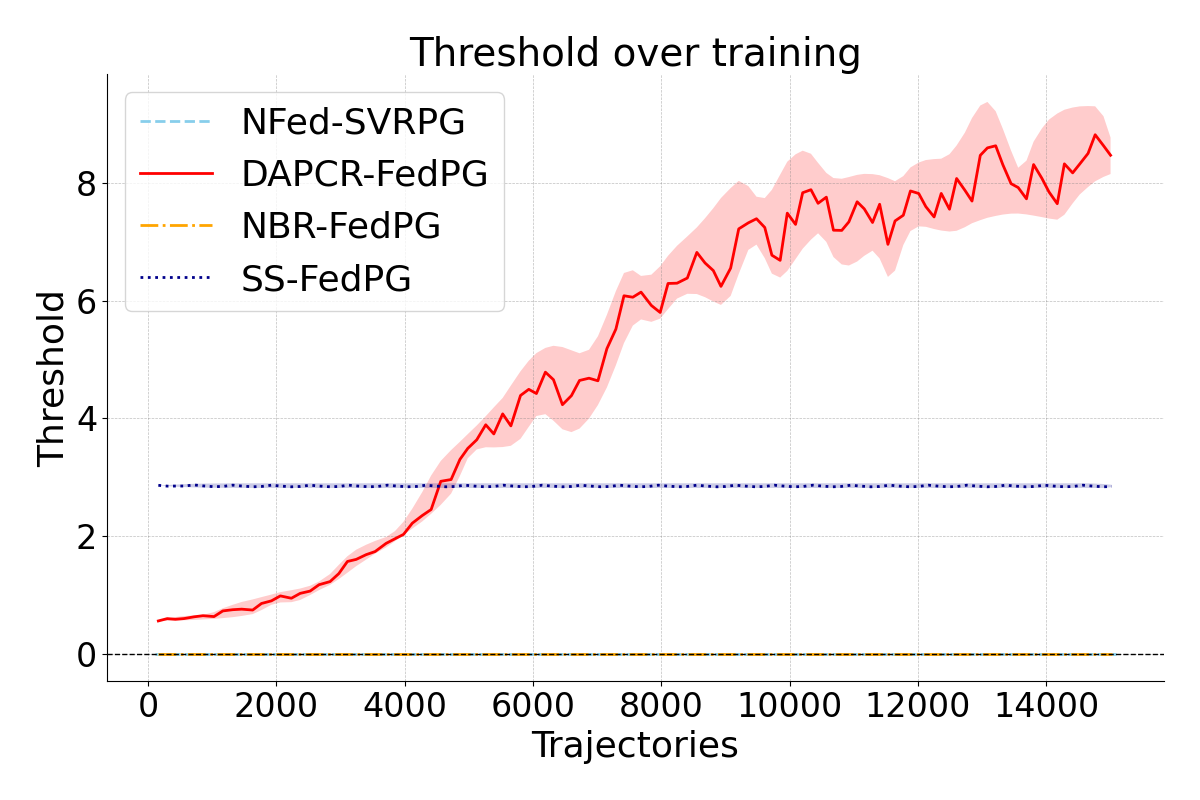}
        \caption{Threshold over training}
        \label{threshold}
    \end{subfigure}
    \hfill
    \begin{subfigure}{0.68\linewidth}
        \centering
        \includegraphics[width=\linewidth]{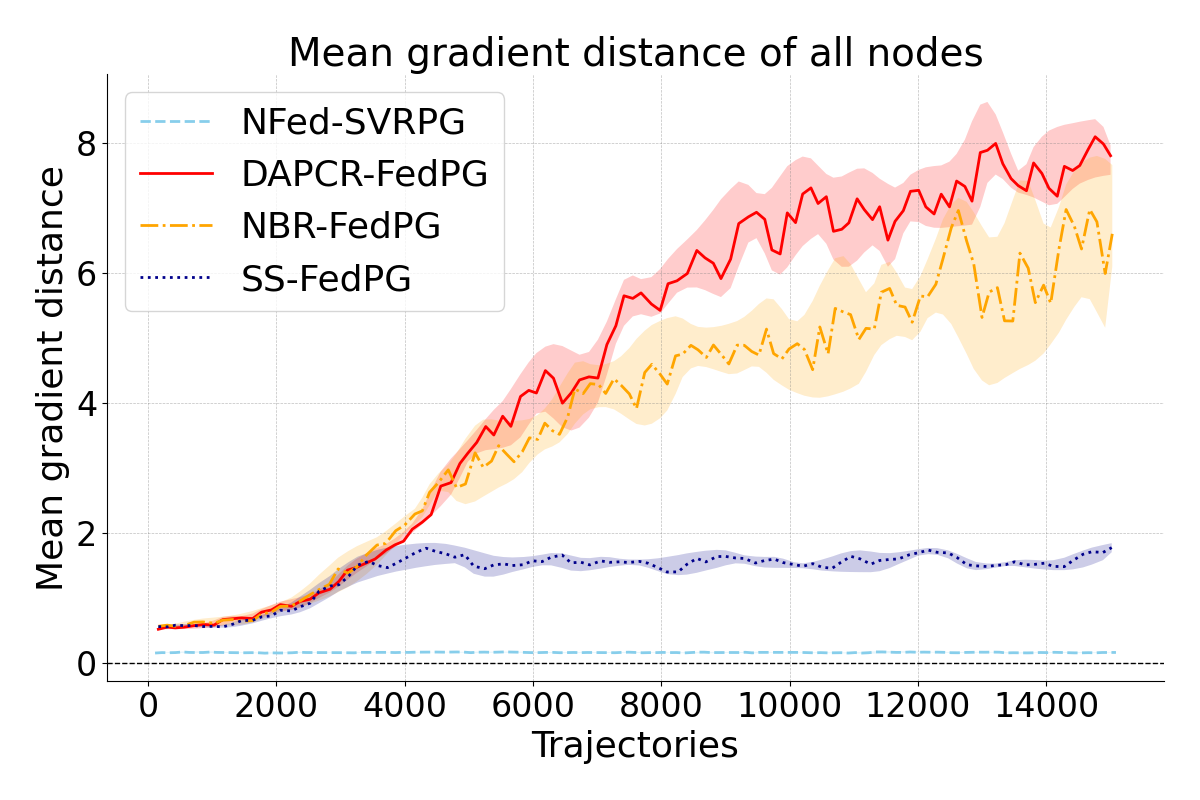}
        \caption{Mean gradient distance of all nodes.}
        \label{grad_norm_mean_all}
    \end{subfigure}
    \caption{Threshold and gradient distance comparison}
    \label{Comp_gradient_d}
\end{figure}

To validate the effectiveness of the proposed DAPCR-FedPG algorithm, a comparison is conducted with the following baseline algorithms under the given settings:
\begin{itemize}
    \item No Byzantine Resilient Federated PG (NBR-FedPG): Local gradients are integrated without filtering \cite{zhuo2019federated}.
    \item Static Screening Federated PG (SS-FedPG): Local gradients are filtered using a fixed static variance bound \cite{fan2021fault}.
    \item No Federated Stochastic Variance-Reduced PG (NFed-SVRPG): Local nodes perform SVRPG individually without FRL \cite{papini2018stochastic}.
\end{itemize}

Fig. \ref{Comp_gradient_d} compares the threshold and gradient distances across the four algorithms. Gradient distance, measured by the Euclidean norm between local gradients, reflects the degree of local policy inconsistency—larger values indicate greater inconsistency but also imply more effective local policy learning. As shown in Fig. \ref{threshold}, NBR-FedPG and NFed-SVRPG maintain zero thresholds, owing to the absence of gradient filtering in the former and the lack of gradient aggregation in the latter. SS-FedPG uses a semi-constant threshold, with fluctuations induced by batch size, offering limited adaptability. In contrast, DAPCR-FedPG dynamically adjusts its threshold in response to increasing gradient distances among local nodes, thereby better accommodating local policy inconsistency. 

Fig. \ref{grad_norm_mean_all} further compares the average gradient distances across all nodes. NFed-SVRPG fails to learn effective policies and remains stagnant. Other algorithms show an initial increase during the first 4,000 trajectories, reflecting effective local learning under FRL. DAPCR-FedPG sustains this trend with a gradual deceleration in growth, indicating both continued refinement and bounded inconsistency. In comparison, NBR-FedPG exhibits slower growth and greater oscillations due to unfiltered noise, while SS-FedPG stagnates early due to its fixed threshold limiting further adaptation.

\begin{figure}[ht]
\centering
    \includegraphics[width=0.68\linewidth]{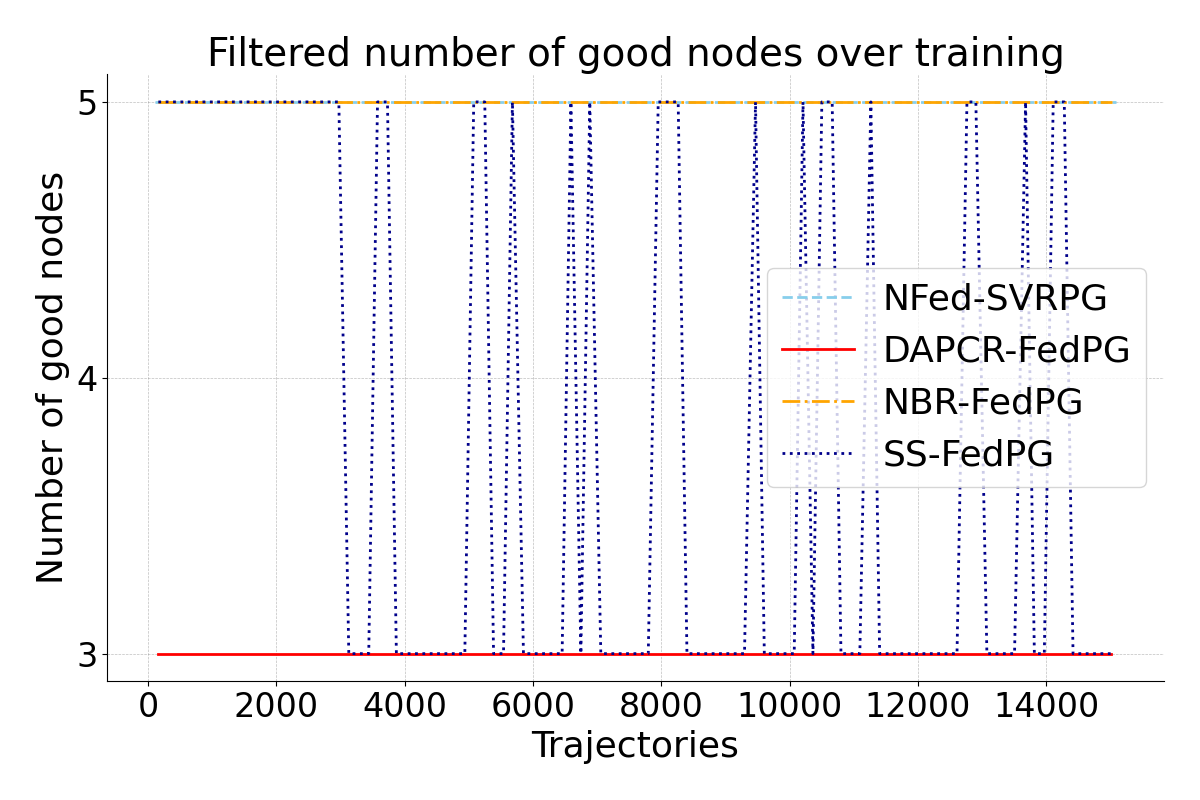}
    \caption{Comparison of good node identification performance.}
    \label{Comp_good}
\end{figure}

Fig. \ref{Comp_good} shows the effectiveness of DTBF by comparing the performance in identifying good nodes. DAPCR-FedPG consistently identifies the correct number of good nodes (3 out of 5 SPs, with 2 being Byzantine). In contrast, NBR-FedPG and NFed-SVRPG lack Byzantine filtering mechanisms, leading to misidentification of all nodes as good. The static threshold in SS-FedPG leads to alternating over-permissiveness and over-restrictiveness, causing unstable filtering—sometimes correctly identifying three good nodes, while at other times misclassifying all nodes as good.

\begin{figure*}[ht]
\centering
\begin{subfigure}[t]{0.325\textwidth}
    \centering
    \includegraphics[width=\linewidth]{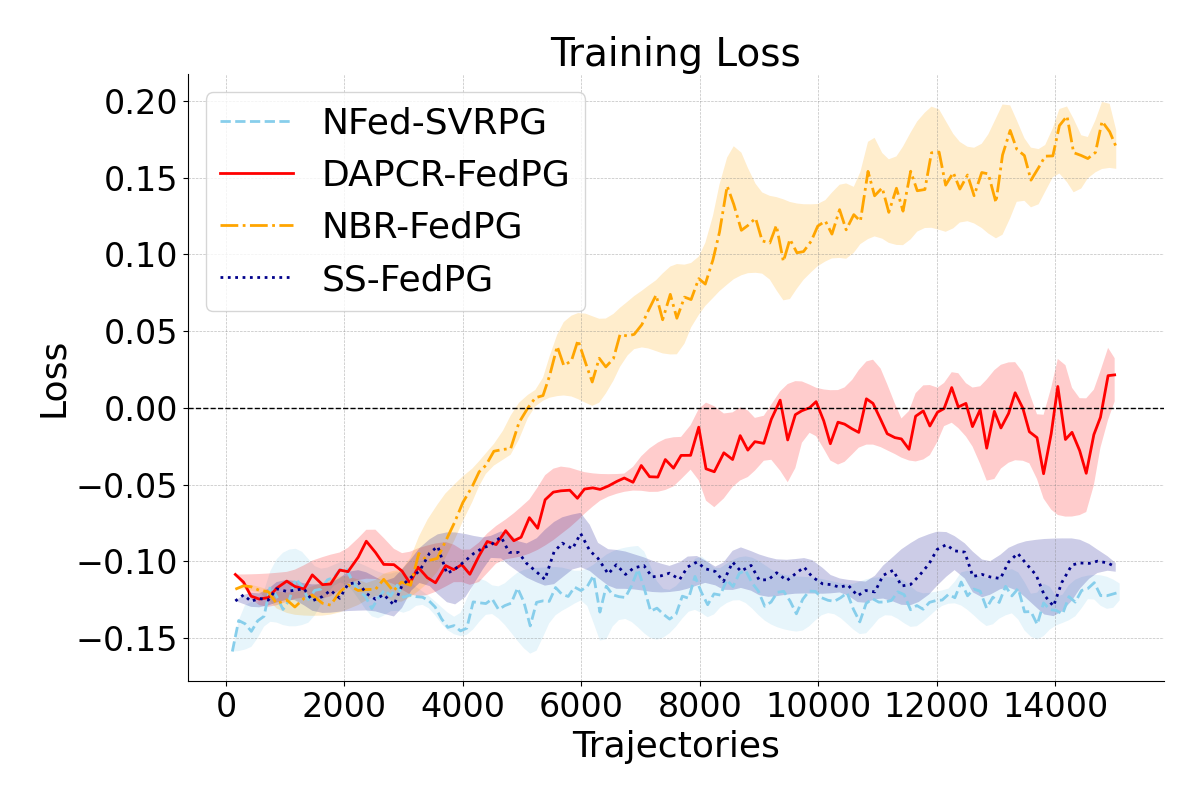}
    \caption{Training loss.}
    \label{t_l}
\end{subfigure}
\begin{subfigure}[t]{0.325\textwidth}
    \centering
    \includegraphics[width=\linewidth]{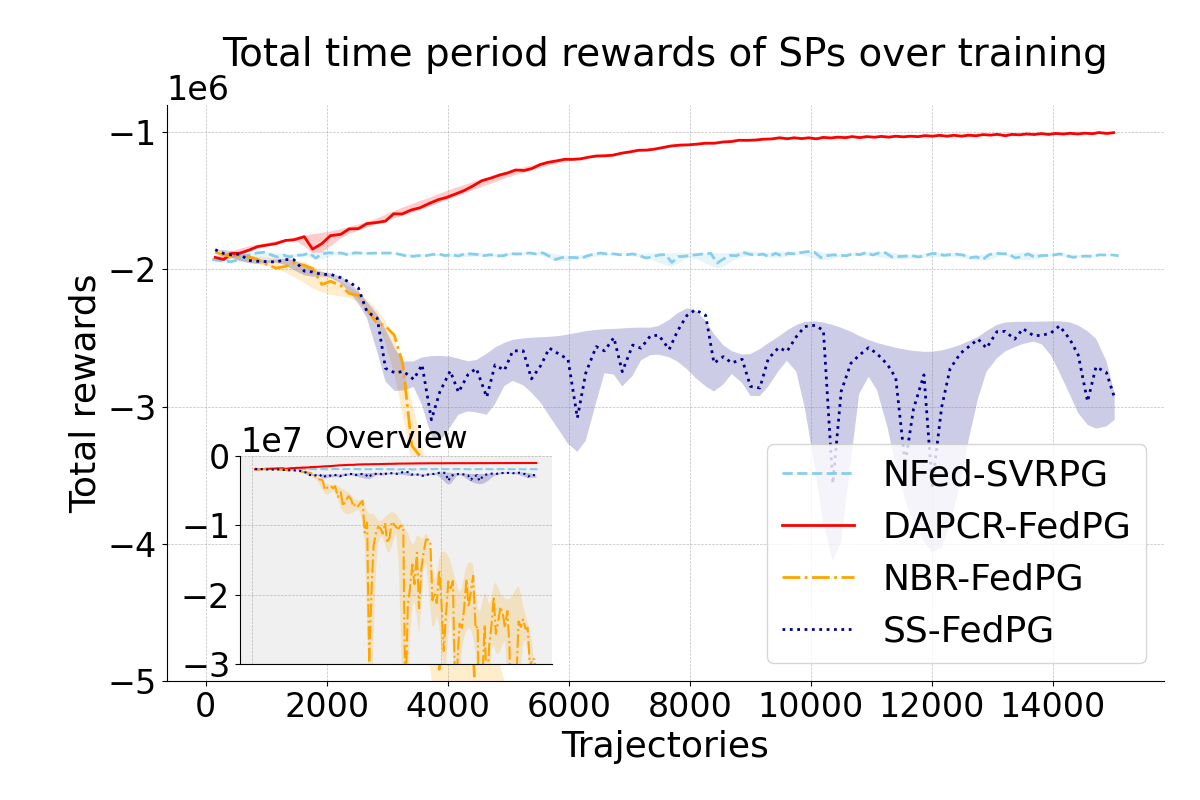}
    \caption{Total rewards of SPs over training.}
    \label{t_r}
\end{subfigure}
\begin{subfigure}[t]{0.325\textwidth}
    \centering
    \includegraphics[width=\linewidth]{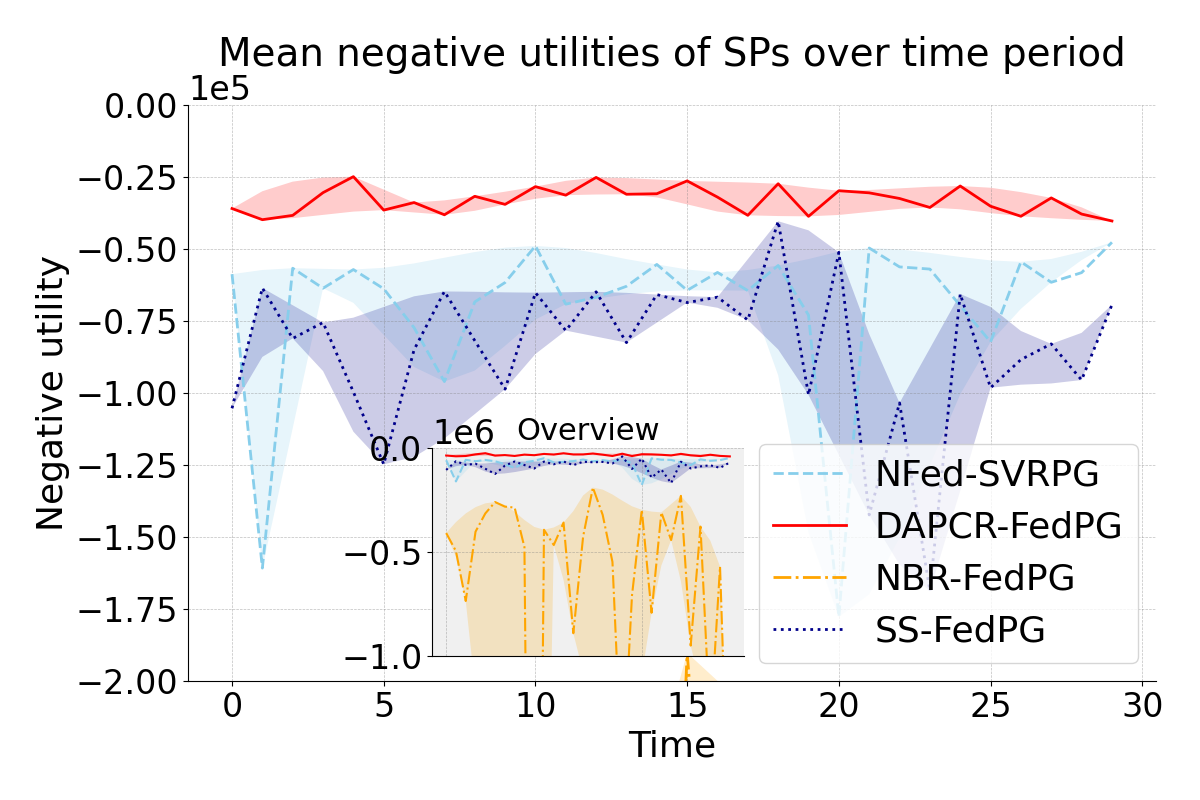}
    \caption{Mean negative utilities of SPs.}
    \label{v_mSPr}
\end{subfigure}
\caption{Comparison of training and validation performance.}
\label{Comp_tv}
\end{figure*}

\begin{figure*}[t]
\centering
\begin{subfigure}[t]{0.325\textwidth}
    \centering
    \includegraphics[width=\linewidth]{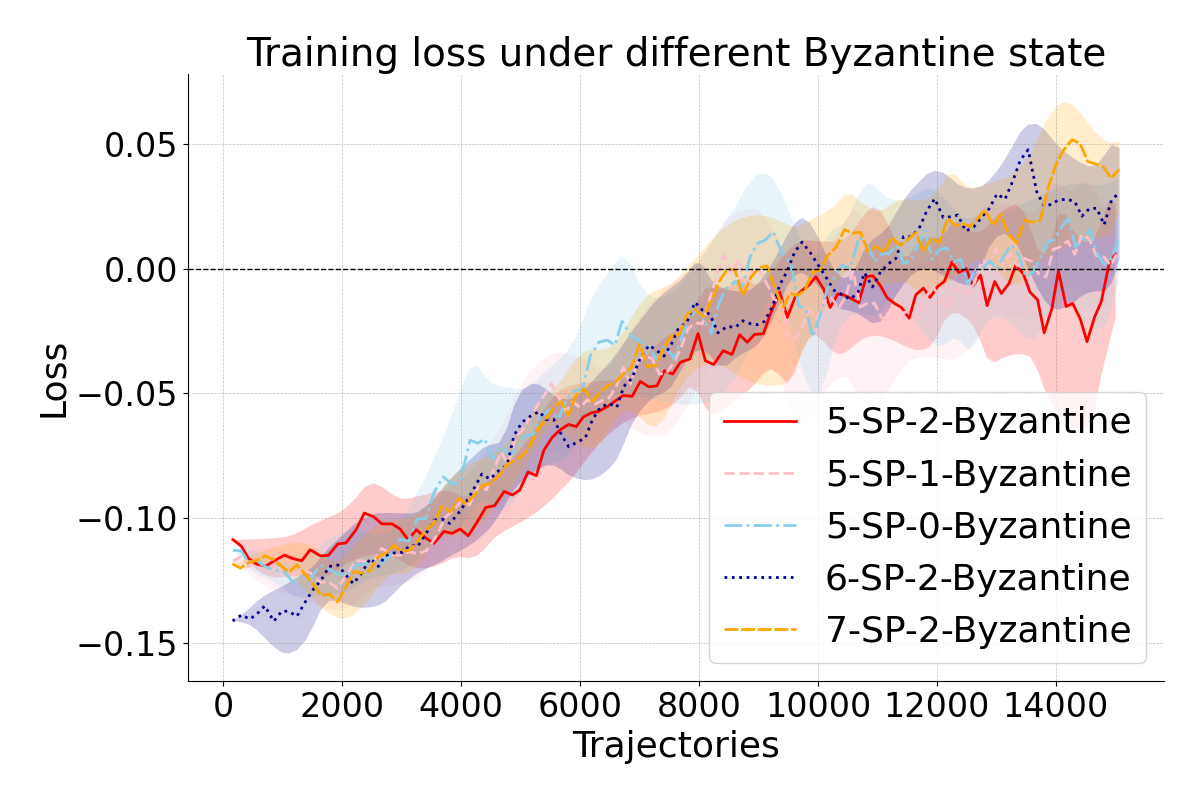}
    \caption{Training loss.}
    \label{p_t_l}
\end{subfigure}
\begin{subfigure}[t]{0.325\textwidth}
    \centering
    \includegraphics[width=\linewidth]{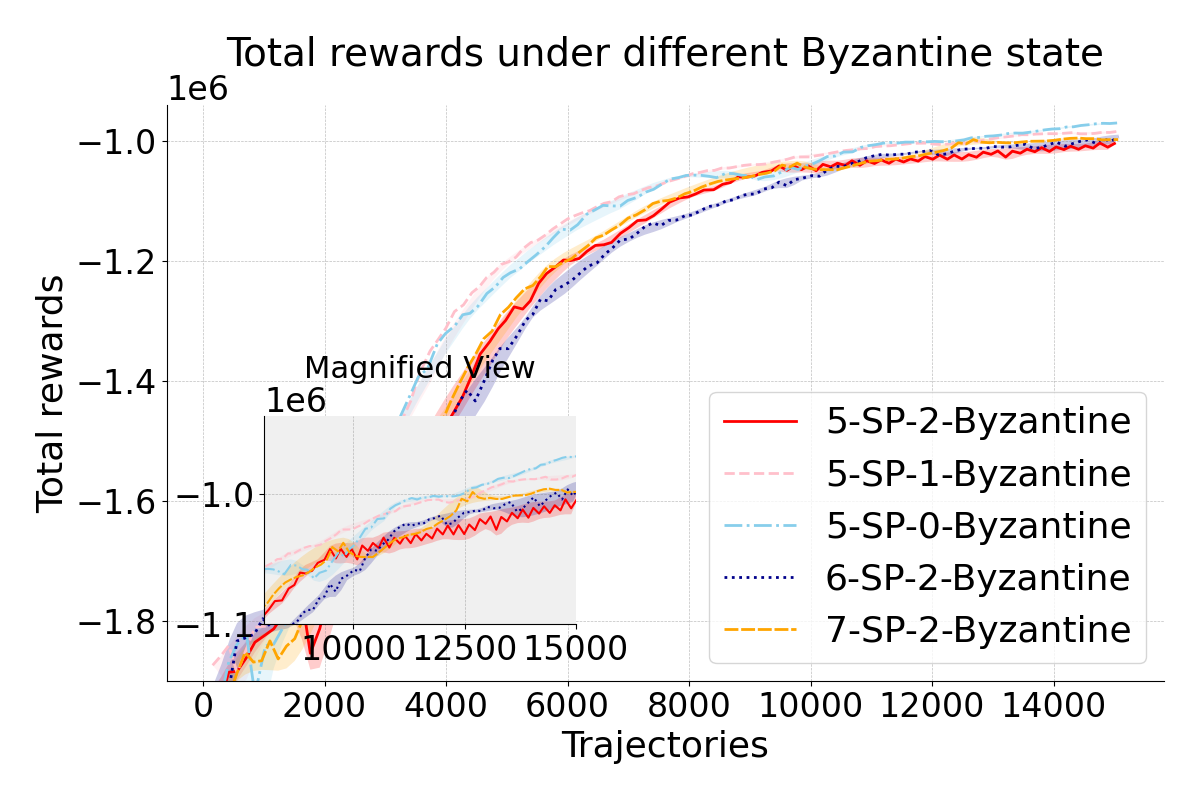}
    \caption{Total rewards of SPs over training.}
    \label{p_t_r}
\end{subfigure}
\begin{subfigure}[t]{0.325\textwidth}
    \centering
    \includegraphics[width=\linewidth]{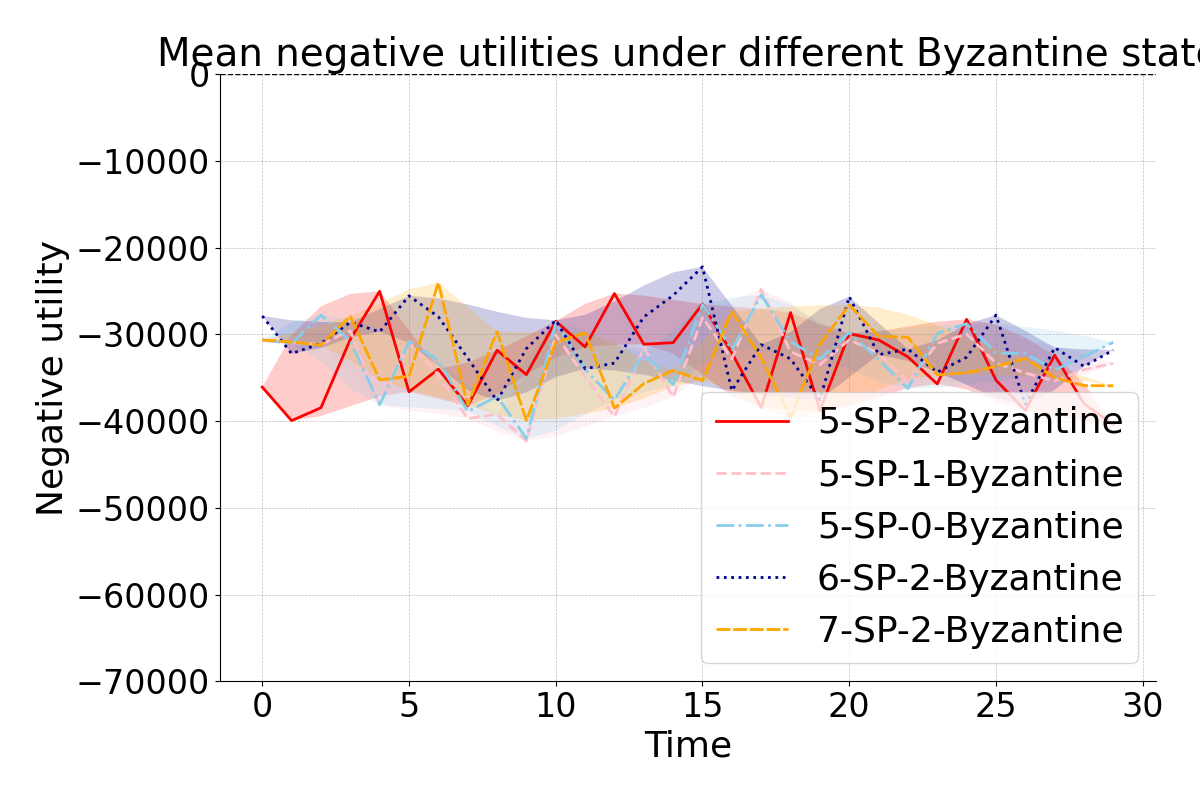}
    \caption{Mean negative utilities of SPs.}
    \label{p_v_mSPr}
\end{subfigure}
\caption{Performance under different Byzantine states.}
\label{para_tv}
\end{figure*}

Fig. \ref{Comp_tv} compares training and validation performance across algorithms. Figs. \ref{t_l} and \ref{t_r} show that DAPCR-FedPG achieves stable convergence, with loss approaching zero and rewards stabilizing around $-1 \times 10^{6}$, indicating effective policy learning. In contrast, NFed-SVRPG, lacking the FRL, exhibits fluctuating loss and rewards around the initialization level, reflecting its failure to learn from local observations alone. SS-FedPG slightly improves loss performance upon NFed-SVRPG but suffers from intermittent filtering failures that degrade reward performance. NBR-FedPG, without fault tolerance, performs the worst, with diverging loss and reward, highlighting the importance of fault tolerance.

Fig. \ref{v_mSPr} presents validation results in terms of mean negative SP utility. DAPCR-FedPG maintains the lowest energy consumption and exhibits superior stability, as reflected in the narrow envelope. This stability is crucial for UAV operations—it ensures predictable energy use, prevents excessive depletion during high-demand periods, enables accurate battery management, and minimizes wear, thereby extending UAV lifespan and reducing maintenance costs. Such robustness is particularly valuable in infrastructure-limited scenarios like disaster recovery and remote communications. In contrast, the baseline methods exhibit higher energy consumption and significantly larger variance envelopes, indicating poorer stability.

\subsection{Parameter Analysis}

To assess the robustness of DAPCR-FedPG, simulation experiments are conducted under varying Byzantine states and network scales. Byzantine states refer to different combinations of SP and Byzantine node counts, while scales reflect varying numbers of hotspots and service types. The base configuration, denoted as 5-SP-2-Byzantine, includes 5 SPs, 2 Byzantine nodes with $\alpha_{\textup{B}}=0.4$, 6 hotspots, and 4 service types. Other configurations are summarized in Table \ref{combined_settings}.

\begin{table}[h]
\centering
\caption{Parameter Settings}
\resizebox{\columnwidth}{!}{%
\begin{tabular}{@{}lccccc|lccccc@{}}
\toprule
\multicolumn{6}{c|}{\textbf{Byzantine States}}                                                                                       & \multicolumn{6}{c}{\textbf{Scales}}                                                                                          \\ \midrule
\textbf{Labels}       & \textbf{$N$} & \textbf{$|\mathcal{N}\setminus\mathcal{Z}|$} & \textbf{$\alpha_{\textup{B}}$} & \textbf{$H$} & \textbf{$K$} & \textbf{Labels}       & \textbf{$N$} & \textbf{$|\mathcal{N}\setminus\mathcal{Z}|$} & \textbf{$\Delta n_{Z}$} & \textbf{$H$} & \textbf{$K$} \\ \midrule
5-SP-1-Byzantine  & 5 & 1 & 0.2   & 6  & 4                & 5-SP-6-service    & 5 & 2 & 24   & 6  & 6 \\ 
5-SP-0-Byzantine  & 5 & 0 & 0     & 6  & 4                & 5-SP-8-service    & 5 & 2 & 48   & 6  & 8 \\ 
6-SP-2-Byzantine  & 6 & 2 & 0.333 & 6  & 4                & 5-SP-8-hotspot    & 5 & 2 & 16   & 8  & 4 \\ 
7-SP-2-Byzantine  & 7 & 2 & 0.286 & 6  & 4                & 5-SP-10-hotspot   & 5 & 2 & 32   & 10 & 4 \\ 
\bottomrule
\end{tabular}}
\label{combined_settings}
\end{table}

\begin{figure*}[t]
\centering
\begin{subfigure}[t]{0.325\textwidth}
    \centering
    \includegraphics[width=\linewidth]{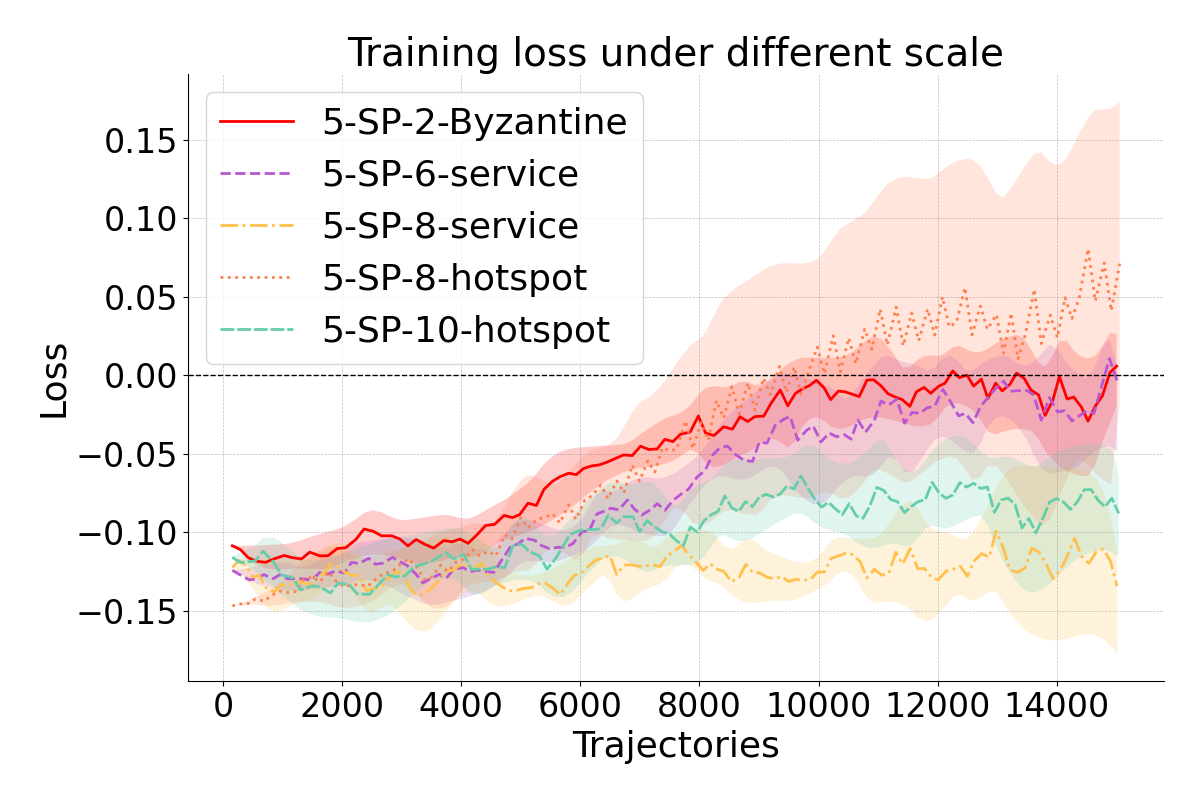}
    \caption{Training loss.}
    \label{p_t_l_s}
\end{subfigure}
\begin{subfigure}[t]{0.325\textwidth}
    \centering
    \includegraphics[width=\linewidth]{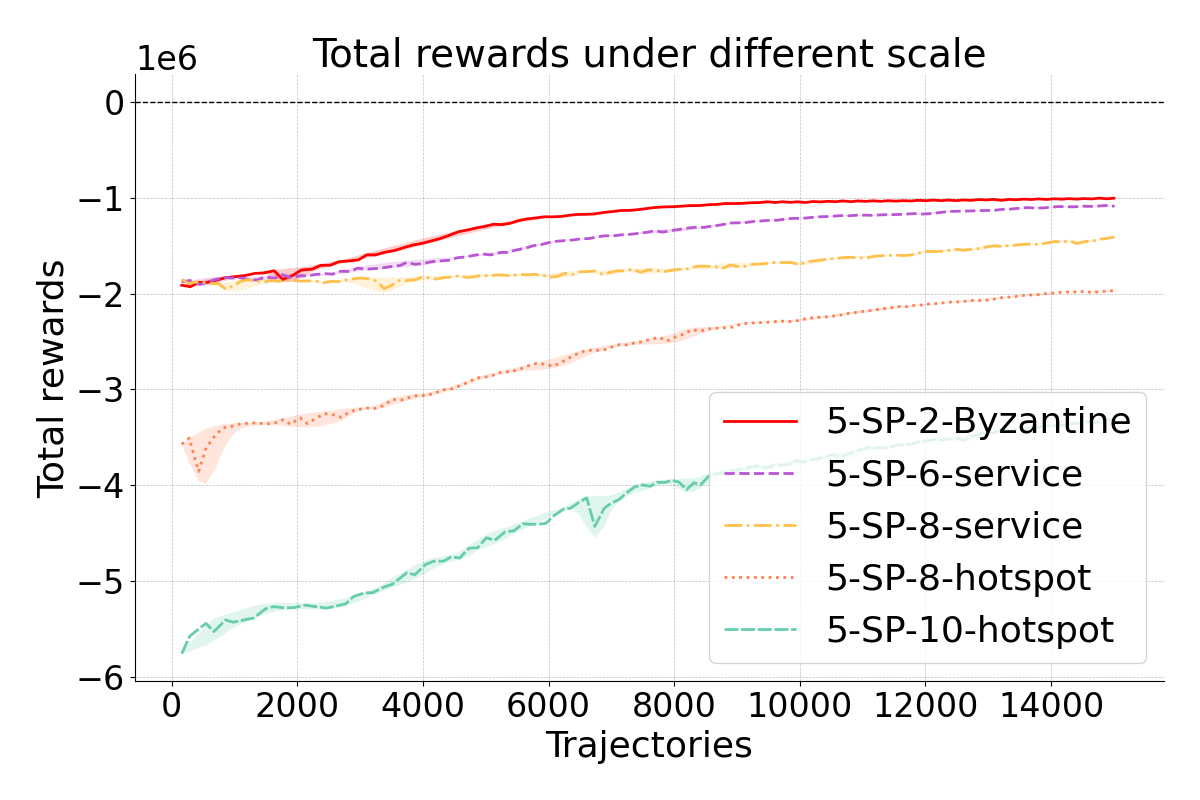}
    \caption{Total rewards of SPs over training.}
    \label{p_t_r_s}
\end{subfigure}
\begin{subfigure}[t]{0.325\textwidth}
    \centering
    \includegraphics[width=\linewidth]{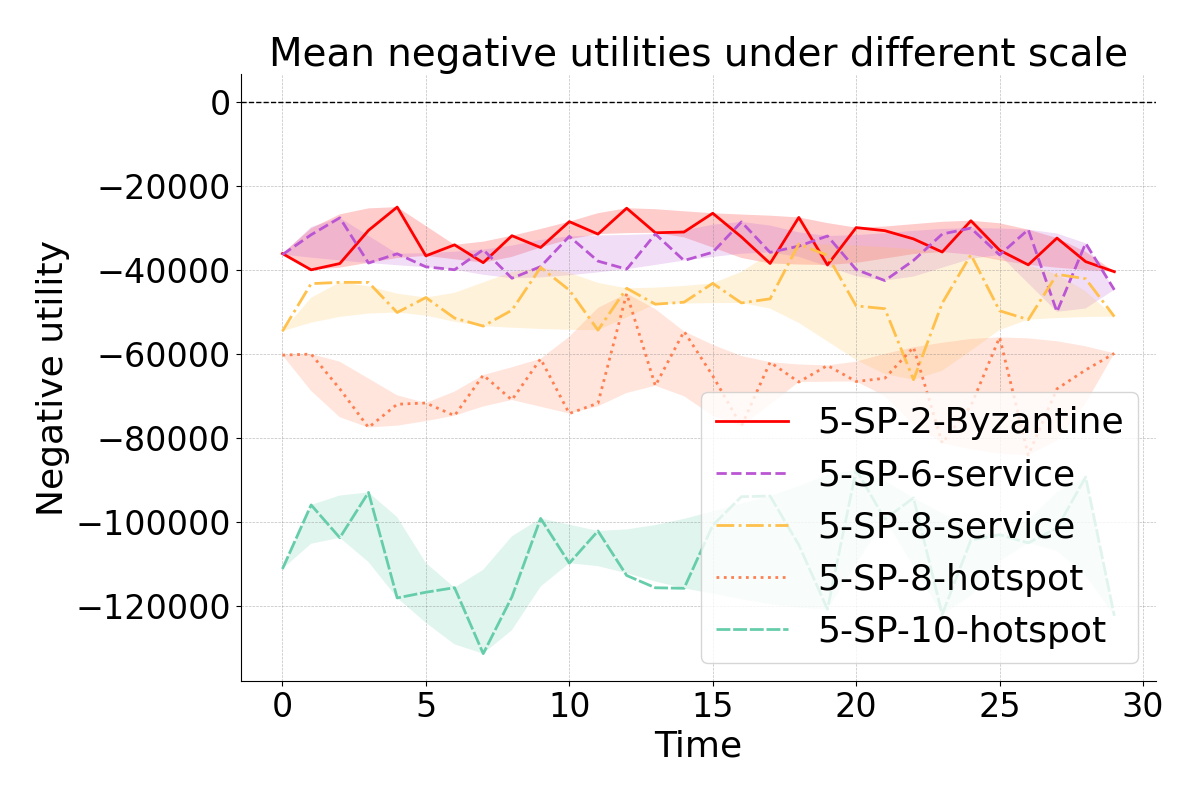}
    \caption{Mean negative utilities of SPs.}
    \label{p_v_mSPr_s}
\end{subfigure}
\caption{Performance under different scales.}
\label{para_tv_s}
\end{figure*}

The training and validation performance under different Byzantine states is presented in Fig. \ref{para_tv}. All states show similar loss performance as shown in Fig. \ref{p_t_l}, confirming the robustness of the DAPCR-FedPG. As seen in Fig. \ref{p_t_r}, smaller Byzantine ratio $\alpha_{\textup{B}}$ (5-SP-0-Byzantine and 5-SP-1-Byzantine) exhibit only slightly faster reward growth, while all states converge to similar rewards. Fig. \ref{p_v_mSPr} further confirms that all settings achieve similar mean negative utility in validation, underscoring the robustness of the DAPCR-FedPG algorithm to Byzantine interference. Moreover, such a performance similarity validates the tolerance of local policy inconsistencies, since the number of SPs increases, interaction complexity and local policy divergence also grow.

Fig. \ref{para_tv_s} presents the performance under different scales, where increasing the number of service types or hotspots expands the action space due to the Dirichlet-based policy structure. Each additional service type increases the action dimension by $2H$, and adding one hotspot increases it by $2K$, since bidding actions are two-dimensional. The resulting increases in action dimension $\Delta n_Z$ are summarized in Table \ref{combined_settings}. As shown in Fig. \ref{p_t_l_s}, 5-SP-6-service and 5-SP-8-hotspot, with relatively small dimensional increases (24 and 16), maintain loss performance close to the base configuration, though 5-SP-8-hotspot exhibits wider fluctuations. In contrast, 5-SP-8-service and 5-SP-10-hotspot show poorer performance due to higher dimensional increases (48 and 32), since a larger action space leads to performance degradation if the policy network architecture remains unchanged.

When total resources remain fixed, increasing the number of hotspots dilutes per-hotspot resource allocation. As shown in Fig. \ref{p_t_r_s} and Fig. \ref{p_v_mSPr_s}, total rewards and utilities for 5-SP-8-hotspot and 5-SP-10-hotspot significantly decrease compared to the base configuration. Although adding service types increases total resources, 5-SP-6-service and 5-SP-8-service still show slight declines in reward and utility due to impaired training effectiveness caused by increased action dimensions.

It is important to note that this performance degradation arises from the unchanged policy network structure rather than the proposed algorithm itself. Despite this structural limitation, all scales still demonstrate convergence in reward and stability in utility, highlighting the robustness of the proposed solution even under increased action space complexity.

\section{Conclusion}\label{conclusion}
This work has presented an effective resource allocation framework for LAE, addressing the challenge of reliable multi-service deployment in infrastructure-limited regions. An auction-based mechanism has been developed to resolve competition among SPs, and the existence of the NE has been established through game-theoretic analysis, thereby guaranteeing the solvability of the long-term utility optimization problem. The proposed resilient FRL algorithm has demonstrated strong fault tolerance and has achieved more efficient integrated resource allocation and improved robustness compared with baseline methods under various parameter settings.

By enabling potential global cooperation among locally self-interested SPs, this work has established a fault-tolerant FRL-based paradigm for competitive LAE systems, which reconciles individual rationality with global efficiency and demonstrates the feasibility of scalable service provisioning in infrastructure-limited environments. Building on this framework, future work will further investigate physically grounded service deployment strategies, including UAV swarm scheduling, user task offloading, and joint satellite–UAV coordination, as well as enhanced robustness against diverse transmission errors and malicious attacks.

\appendices
\section{The Proof of Theorem \ref{AUTH_AUC}}\label{appA}
\renewcommand{\theequation}{A.\arabic{equation}}
\setcounter{equation}{0}
This theorem is proved by contradiction. Assume an SP obtains higher utility through false bidding. Denote this false bid as $(F_{nhk}^{\textup{F}}(t),B_{nhk}^{\textup{F}}(t))$, which corresponds to a committed delay $\hat{T}_{nhk}(t)$). The SP's actual provision is $(F_{nhk}^{\textup{A}}(t),B_{nhk}^{\textup{A}}(t))$, which commits to an actual delay $T_{nhk}(t)$). Here,
$F_{nhk}^{\textup{F}}(t) > F_{nhk}^{\textup{A}}(t)$ and $B_{nhk}^{\textup{F}}(t) > B_{nhk}^{\textup{A}}(t)$ hold. Let $U_{nhk}^{\textup{F}}(t)$ and $U_{nhk}^{\textup{A}}(t)$ represent the utilities under false and authentic bids on service type $k$ at hotspot $h$, respectively. Based on the assumption above, there exist three cases: 

Case 1) SP $n$ still loses the auction. According to Eq. \eqref{EC_SP}, its utility remains fixed at $U_{nhk}^{\textup{F}}(t) = -C_{nk}T_{nhk}(t)=U_{nhk}^{\textup{A}}(t)$, due to the fact $W_{nhk}(t)=0$. 

Case 2) SP $n$ wins the auction from losing. As shown in Eqs. \eqref{T_comp} and \eqref{T_comm}, lower values of $F_{nhk}(t)$ and $B_{nhk}(t)$ increase computation and communication delays, thereby increasing processing delay. Therefore, $\hat{T}_{nhk}(t)<T_{nhk}$, leading to $\hat{W}_{nhk}(t)=0$ and $U_{nhk}^{\textup{F}}(t) = - C_{nk}T_{nhk}(t) = U_{nhk}^{\textup{A}}(t)$.

Case 3) SP $n$ still wins the auction. For authentic bidding, the committed and actual delays are the same, leading to $\hat{W}_{nhk}(t)=1$. Therefore, $U_{nhk}^{\textup{A}}(t) = - C_{nk}T_{nhk}(t) + V_{nk}$. For false bidding, $\hat{W}_{nhk}(t)=0$, leading to $U_{nhk}^{\textup{F}}(t) = - C_{nk}T_{nhk}(t) < U_{nhk}^{\textup{A}}(t)$.

Combining the above cases, we have:
\begin{equation}
    U_{nhk}^{\textup{F}}(t) \leq U_{nhk}^{\textup{A}}(t),
\end{equation}
which indicates that false bids lead to lower utility, contradicting the assumption. This thus proves Theorem \ref{AUTH_AUC}.

\section{The Proof of Theorem \ref{Theo:equivalent}}\label{appE}
\renewcommand{\theequation}{B.\arabic{equation}}

Rewrite the utilities as $U_{n}^{\textup{SP}}(t) = U_{n}^{\textup{SP}}(a_{n}(t), \mathbf{a}_{-n}(t))$ and $\bar{U}_{n}^{\textup{SP}}(t) = \bar{U}_{n}^{\textup{SP}}(a_{n}(t), \mathbf{a}_{-n}(t))$. For any $a_{n}(t), a_{n}'(t)$, assume $T_{nhk}(t) \ge T_{nhk}'(t)$, i.e., $a_{n}'(t)$ is the better action. Then, three cases arise:

Case 1) from $a_{n}(t)$ to $a_{n}'(t)$, SP $n$ wins the auction of hotspot $h$ on service $k$. Therefore, we have:

    \begin{equation}
    \begin{split}
        & U_{n}^{\textup{SP}}(a_{n}(t),\mathbf{a}_{-n}(t))=-C_{nk}T_{nhk}(t) < \\
        & U_{n}^{\textup{SP}}(a_{n}'(t),\mathbf{a}_{-n}(t))=-C_{nk}T_{nhk}'(t)+V_{nk},
    \end{split}
    \end{equation}
    and,
        \begin{equation}
    \begin{split}
        & \bar{U}_{n}^{\textup{SP}}(a_{n}(t),\mathbf{a}_{-n}(t))=-C_{nk}T^{\textup{Pen}} < \\
        & \bar{U}_{n}^{\textup{SP}}(a_{n}'(t),\mathbf{a}_{-n}(t))=-C_{nk}T_{nhk}'(t).
    \end{split}
    \end{equation}

Case 2) from $a_{n}(t)$ to $a_{n}^{'}(t)$, the auction of hotspot $h$ on service $k$ remains winning. Therefore, we have:

    \begin{equation}
    \begin{split}
        & U_{n}^{\textup{SP}}(a_{n}(t),\mathbf{a}_{-n}(t))=-C_{nk}T_{nhk}(t)+V_{nk} \leq \\
        & U_{n}^{\textup{SP}}(a_{n}'(t),\mathbf{a}_{-n}(t))=-C_{nk}T_{nhk}'(t)+V_{nk},
    \end{split}
    \end{equation}
and,
        \begin{equation}
    \begin{split}
        & \bar{U}_{n}^{\textup{SP}}(a_{n}(t),\mathbf{a}_{-n}(t))=-C_{nk}T_{nhk}(t) \leq \\
        & \bar{U}_{n}^{\textup{SP}}(a_{n}'(t),\mathbf{a}_{-n}(t))=-C_{nk}T_{nhk}'(t).
    \end{split}
    \end{equation}

Case 3) from $a_{n}(t)$ to $a_{n}^{'}(t)$, the auction of hotspot $h$ on service $k$ remains losing. Therefore, we have:
    \begin{equation}
    \begin{split}
        & U_{n}^{\textup{SP}}(a_{n}(t),\mathbf{a}_{-n}(t))=-C_{nk}T_{nhk}(t) \leq \\
        & U_{n}^{\textup{SP}}(a_{n}'(t),\mathbf{a}_{-n}(t))=-C_{nk}T_{nhk}'(t),
    \end{split}
    \end{equation}
and,
        \begin{equation}
    \begin{split}
        & \bar{U}_{n}^{\textup{SP}}(a_{n}(t),\mathbf{a}_{-n}(t))=-C_{nk}T^{\textup{Pen}} =\\
        & \bar{U}_{n}^{\textup{SP}}(a_{n}'(t),\mathbf{a}_{-n}(t))=-C_{nk}T^{\textup{Pen}}.
    \end{split}
    \end{equation}

    The following inequality set is always satisfied across the aforementioned cases:
    \begin{equation}
        \left\{\begin{matrix}
        U_{n}^{\textup{SP}}(a_{n}(t),\mathbf{a}_{-n}(t)) \leq U_{n}^{\textup{SP}}(a_{n}'(t),\mathbf{a}_{-n}(t)),\\
        \bar{U}_{n}^{\textup{SP}}(a_{n}(t),\mathbf{a}_{-n}(t)) \leq \bar{U}_{n}^{\textup{SP}}(a_{n}'(t),\mathbf{a}_{-n}(t)),
\end{matrix}\right.
    \end{equation}
which means that the utility functions defined by $U_{n}^{\textup{SP}}(t)$ and $\bar{U}_{n}^{\textup{SP}}(t)$ are qualitatively equivalent and can effectively reflect the rewards associated with actions in quantitative analysis. Therefore, Theorem \ref{Theo:equivalent} is proven.

\section{The Proof of Theorem \ref{at_least_1NE}}\label{appB}
\renewcommand{\theequation}{C.\arabic{equation}}

    The potential function can be rewritten as $$\Phi^{\textup{SP}}(t) = \Phi^{\textup{SP}}(a_{n}(t),\textbf{a}_{-n}(t)).$$
    Assume $a_{n}^{*}(t)\ne a_{n}(t)$, there are three cases to consider:
    
    Case 1) from $a_{n}(t)$ to $a_{n}^{*}(t)$, SP $n$ wins the auction of hotspot $h$ on service $k$. Therefore, we have:
    \begin{equation}
    \begin{split}
        &\bar{U}_{n}^{\textup{SP}}(a_{n}^{*}(t),\textbf{a}_{-n}(t)) - \bar{U}_{n}^{\textup{SP}}(a_{n}(t),\textbf{a}_{-n}(t))\\
        = & C_{nk}[T^{\textup{Pen}}-T_{nhk}(t)],
    \end{split}
    \end{equation}
    \begin{equation}
    \begin{split}
        &\Phi^{\textup{SP}}(a_{n}^{*}(t),\textbf{a}_{-n}(t)) - \Phi^{\textup{SP}}(a_{n}(t),\textbf{a}_{-n}(t))\\
        = & - [W_{nhk}(a_{n}^{*}(t),\textbf{a}_{-n}(t))T_{nhk}(t)+ \\ 
        &(1-W_{nhk}(a_{n}^{*}(t),\textbf{a}_{-n}(t)))T^{\textup{Pen}}] +\\
        & [W_{nhk}(a_{n}(t),\textbf{a}_{-n}(t))T_{nhk}(t)+ \\
        & (1-W_{nhk}(a_{n}(t),\textbf{a}_{-n}(t)))T^{\textup{Pen}}] \\
        = & T^{\textup{Pen}} - T_{nhk}(t).
    \end{split}
    \end{equation}
    
    Case 2) from $a_{n}(t)$ to $a_{n}^{*}(t)$, the auction results remain still. This case can be further divided into two subcases:

    Case 2.1) the auction remains losing. Therefore, we have:
        \begin{equation}
        \bar{U}_{n}^{\textup{SP}}(a_{n}^{*}(t),\textbf{a}_{-n}(t)) = \bar{U}_{n}^{\textup{SP}}(a_{n}(t),\textbf{a}_{-n}(t)),
    \end{equation}
    \begin{equation}
        \Phi^{\textup{SP}}(a_{n}^{*}(t),\textbf{a}_{-n}(t)) = \Phi^{\textup{SP}}(a_{n}(t),\textbf{a}_{-n}(t)).
    \end{equation}
    
    Case 2.2) the auction remains winning. Therefore, we have:
    \begin{equation}
        \bar{U}_{n}^{\textup{SP}}(a_{n}^{*}(t),\textbf{a}_{-n}(t)) - \bar{U}_{n}^{\textup{SP}}(a_{n}(t),\textbf{a}_{-n}(t))=  C_{nk}\Delta T_{nhk}(t),
    \end{equation}
    \begin{equation}
        \Phi^{\textup{SP}}(a_{n}^{*}(t),\textbf{a}_{-n}(t)) - \Phi^{\textup{SP}}(a_{n}(t),\textbf{a}_{-n}(t)) = \Delta T_{nhk}(t),
    \end{equation}
    where $\Delta T_{nhk}(t) = T_{nhk}(a_{n}(t)) - T_{nhk}(a_{n}^{*}(t))$.

    Case 3) from $a_{n}(t)$ to $a_{n}^{*}(t)$, SP $n$ loses the auction of hotspot $h$ on service $k$. Therefore, we have:
    \begin{equation}
    \begin{split}
        &\bar{U}_{n}^{\textup{SP}}(a_{n}^{*}(t),\textbf{a}_{-n}(t)) - \bar{U}_{n}^{\textup{SP}}(a_{n}(t),\textbf{a}_{-n}(t))\\
        = & C_{nk}[T_{nhk}(t) - T^{\textup{Pen}}],
    \end{split}
    \end{equation}
    \begin{equation}
    \begin{split}
        &\Phi^{\textup{SP}}(a_{n}^{*}(t),\textbf{a}_{-n}(t)) - \Phi^{\textup{SP}}(a_{n}(t),\textbf{a}_{-n}(t))\\
        = & T_{nhk}(t) - T^{\textup{Pen}}.
    \end{split}
    \end{equation}

    Combining all three cases, the following equality holds:
\begin{equation}\label{w_p_f}
    \begin{split}
         & \bar{U}_{n}^{\textup{SP}}(a_{n}^{*}(t),\textbf{a}_{-n}(t)) - \bar{U}_{n}^{\textup{SP}}(a_{n}(t),\textbf{a}_{-n}(t)) \\
         = & C_{nk}[\Phi^{\textup{SP}}(a_{n}^{*}(t),\textbf{a}_{-n}(t)) - \Phi^{\textup{SP}}(a_{n}(t),\textbf{a}_{-n}(t))].
    \end{split}
    \end{equation}
    
    $C_{nk}$ can be considered constant across services $\forall k\in \mathcal{K}$, since UAVs deployed by the same SP typically share similar parameters. Therefore, any bidding-induced change in auction results constitutes a composition of the three cases above, ensuring that Eq. \eqref{w_p_f} always holds. Accordingly, the stage game $\mathcal{G}_{t}^{\textup{SP}}$ qualifies as a potential game with at least one NE, which proves Theorem \ref{at_least_1NE}.

    \section{The Proof of Theorem \ref{P1_solution}}\label{appC}

    The theorem is proved by contradiction. Assume the solution ${\hat{\textbf{a}}(t)}_{t \in \mathcal{T}}$ to the proposed problem yields higher utility than the NE set ${\textbf{a}^{*}(t)}_{t \in \mathcal{T}}$. According to Definition \ref{defi_NE}, each SP would then benefit from unilaterally deviating from $a_{n}^{*}(t)$ to $\hat{a}_{n}(t)$, violating the NE condition. Hence, the NE set maximizes each SP’s utility, completing the proof of Theorem \ref{P1_solution}.

\bibliographystyle{IEEEtran}
    \bibliography{mylib}

\end{document}